\documentclass{emulateapj}

\usepackage{graphicx}
\usepackage{natbib}
\usepackage{amsmath}
\usepackage{dsfont}
\usepackage{color}
\newcommand\red{\textcolor{black}}
\newcommand\redd{\textcolor{black}}

\newcommand\microk{$\mu$K}

\definecolor{ForestGreen}{rgb}{0.3,0.7,0.3}

\newcommand\zobov{{\scshape zobov}}

\def\apjl{{ApJ}}                
\def\apjs{{ApJS}}               
\def\ao{{Appl.~Opt.}}           
\def\aap{{A\&A}}                

\begin{document}
 
\title{A Possible Cold Imprint of Voids on the Microwave Background Radiation} 

\author{Yan-Chuan Cai$^{1}$\thanks{E-mail: y.c.cai@durham.ac.uk}, Mark C. Neyrinck$^{2}$, Istv\'{a}n Szapudi$^{3}$, Shaun Cole$^{1}$ and Carlos S. Frenk$^{1}$} 
\affil{
 $^1$Institute for Computational Cosmology, Department of Physics, University of Durham, South Road, Durham DH1 3LE, UK \\
 $^2$Department of Physics and Astronomy, The Johns Hopkins University, 3701 San Martin Drive, Baltimore, MD 21218, USA\\
 $^3$Institute for Astronomy, University of Hawaii, 2680 Woodlawn Drive, Honolulu HI 96822, USA}

\begin{abstract}
We measure the average temperature decrement on the cosmic microwave background (CMB) produced by voids selected in the SDSS DR7
spectroscopic redshift galaxy catalog, spanning redshifts $0<z<0.44$. We find an imprint of amplitude between $2.6$ 
and $2.9 $\microk  as viewed through a compensated top-hat filter scaled to the radius of each void; we assess the statistical significance of the imprint 
at $\sim 2\sigma$. We make crucial use of $N$-body simulations to calibrate our analysis. As expected, we find that large voids produce 
cold spots on the CMB through the Integrated Sachs-Wolfe (ISW) effect. However, we also find that small voids in the halo density field 
produce hot spots, because they reside in contracting, larger-scale overdense regions.
This is an important effect to consider when stacking CMB imprints from voids of different radius. 
We have found that the same filter radius that gives the largest ISW signal in simulations also yields 
close to the largest detected signal in the observations. However, although it is low in significance, our measured 
signal is much higher-amplitude than expected from ISW in the concordance $\Lambda$CDM universe. The discrepancy is 
also at the $\sim$2$\sigma$ level. We have demonstrated that our result is robust against the varying of 
thresholds over a wide range. 
\end{abstract}

\keywords{cosmic background radiation --- cosmology: observations ---
large-scale structure of universe --- methods: statistical}

\section{introduction}
The late-time Integrated Sachs-Wolfe effect (ISW) \citep{Sachs67} is direct evidence of cosmic acceleration \citep{CrittendenTurok96}.
However, detection of the ISW effect by the cross-correlation of large-scale structure and the cosmic microwave background (CMB) is challenging, due to 
large cosmic variance and possible systematics. Combined analysis of a few different galaxy/QSO surveys 
has yielded a signal of estimated significance $4\sigma$ \citep{Giannatonio08,Ho08, Giannatonio12}, although there are alternative views \citep[e.g., ][]{Rassat2007, Sawangwit10,Bielby2010,Lopez-Corredoira2010,HM2010}. These, and most other, analyses have used a cross-correlation-function method. 

Another method exploits the physical insight that in the presence of dark energy, linear-scale voids and superclusters stretch faster than they can grow through gravity, and thus produce cold and hot spots respectively in the CMB. \citet[][G08]{Granett08} stacked 100 such quasilinear-scale structures, reporting a detection of corresponding cold and hot spots at {4-$\sigma$} significance. Even without dark energy, the potential can change on nonlinear scales, producing CMB imprints; this is known as the Rees-Sciama (RS) 
effect \citep{Rees68}. We refer to the full effect as ISWRS. The nonlinear RS effect may confuse an interpretation of an ISW-like detection as a signal of dark energy, but it
is expected only at the $\sim 10\%$ level at $z<1$ on sub-degree scales \citep{Cai09, Smith2009, Cai10}.  
The significance level of the G08 detection from a single galaxy sample seems to be higher than the cross-correlation method, 
and the amplitude of the signal is found to be 2 to 3$\sigma$ higher than estimates from simulations; this indicates tension at the 
with the concordance cosmology at $z \sim 0.5$ \citep[G08,][]{Nadathur12,PapaiSzapudi2010, PapaiEtal2011}.
 Incorporating the contribution of 
non-linearity seems unable to reduce this tension \citep{Nadathur2012, Flender12, HM12}.  It is therefore important to check if such a signal/tension persists 
at other epochs of the Universe, and in other void catalogs. 

In this paper, we re-investigate this issue using a new, independent SDSS void catalog which uses the \zobov\ \citep{Neyrinck08} void-finder, as G08 used\red{\footnote{The void catalogue used in this paper can be obtained at $\url{\rm http://}$skysrv.pha.jhu.edu$/\sim$mneyrinck$/$DR7voids.tgz}}. Compared to G08, which used photometric redshifts, the current sample uses galaxies both with accurate spectroscopic redshifts, and, at low redshift, much higher sampling, allowing more accurate knowledge of each void's physical structure.  The current sample covers the redshift range $0<z<0.44$, complementing to the previous catalog, which spans $0.4<z<0.75$. 

Voids identified in a galaxy density field do not necessarily pick out the optimal structures for ISW detection. First, discrete sampling of the density field 
will lead to spurious voids, from Poisson noise. Second, voids of the same density contrast and size may reside in different large-scale environments, changing their 
ISW signals. \red{Pruning void catalogues is important for optimizing the ISW detection, but it is crucial to do so {\it a priori} (e.g.\ theoretically), rather than  {\it a posteriori}, (e.g.\ claiming the largest detection over some ad hoc parameter). Indeed, using simulations, we are able to clean the void catalogues based on physically motivated reasons {\it prior} to inspecting the results. This is an important step to avoid {\it a posteriori} bias in this type of analysis.}


We appreciate that similar void catalogues constructed by \citet{Sutter12} have been used independently 
for ISW detection \citep{Ilic, Planck}. Thus we do not intend to repeat the same analysis using exactly the same catalogues. Rather, we will use our own version of void catalogues, which have subtle differences with that of \citet{Sutter12}. Very recently, \citet{Nadathur2013} 
produced another void catalog from the same volume limited galaxy samples. While all these catalogues may differ from each other, it is unclear 
which is optimal for ISW detection. We discuss our results in the context of these recent measurements in the conclusion. In the present paper, we concentrated on aligning the void detection algorithm as closely as possible between our simulations and the observations. 

In Section 2 of our paper, we briefly describe the void catalog, and details of the stacking procedure. Section 3 presents simulations of the void catalog and the ISWRS signal. The main results are presented in Section 4, including systematic tests. Conclusions and discussion are presented in Section 5.

\section{Void catalog}
 
For our analysis, we use a set of voids detected from the Sloan Digital Sky Survey Data Release 7 (SDSS DR7) main-galaxy (MGS) \red{\citep{Strauss2002}}, and luminous-red-galaxy (LRG) samples \red{\citep{Eisenstein2001}}, covering 8500 deg$^2$ on the sky. These are the same galaxy catalogs as used for void finding by \citet[][S12]{Sutter12}. The MGS has a redshift range of $0<z<0.2$ and the LRG sample a range $0.16<z<0.44$. Six volume-limited samples are made out of these two samples. They are $dim1$ ($0<z<0.05$), $dim2$ ($0.05<z<0.1$), $bright1$ ($0.1<z<0.15$), $bright2$ ($0.15<z<0.2$), $lrgdim$ ($0.16<z<0.36$) and $lrgbright$ ($0.36<z<0.44$). The number densities decrease with increasing sample redshift.

We use the original \zobov\ \citep{NeyrinckEtal2005,Neyrinck08} algorithm to find voids, both in the simulations and the observed samples. In the \zobov\ paper \citep{Neyrinck08}, \zobov\ is called `parameter-free' -- indeed, it can return a parameter-free set of voids and subvoids, nested catchment basins around local density minima as detected in a watershed transform \citep[e.g.][]{PlatenEtal2007}. \red{A parameter-free Voronoi tessellation \citep[e.g.][]{SchaapVdw2000} is used to measure each galaxy's density, and set of neighbours.}  

In \zobov's parameter-free mode, voids can far exceed what might be considered their most physically meaningful extent, since the largest void detected will encompass the entire survey, the density catchment around the globally minimum-density galaxy. So, we use a criterion to halt the joining of `zones' together to form voids.  \red{A `zone' is a subvoid, i.e.\ a density catchment, at the bottom of which is a local density minimum.  Dividing zones from each other are density ridges, where density gradients head down to different density minima on either side. A zone $z$ is not added to a void $v$ if the lowest-density galaxy on the density ridge between $v$ and $z$ exceeds $0.2$}. This threshold value is the fiducial density of a top-hat void after shells cross on its edge, in the spherical expansion model \citep[e.g.][]{ShethVDWeygaert2004}. This step does not affect the number of voids (which corresponds exactly to the number of local density minima), but does affect total void volumes, which are used in our measurement.

There end up being some slight differences between our catalog and the \citet{Sutter12} catalogs. We do not track down every difference, but we list here several slight differences in the implementation.  \citet{Sutter12} explicitly use hierarchical information in splitting voids from each other, which may result in slight differences in how voids are split at the edges \citep{SutterEtal2013b}. They also eliminate voids with
overdensities $\delta > -0.8$, as estimated in a top-hat sphere of radius $(1/4)r_{\rm v}$ about the volume-weighted void centre, where $r_{\rm v}$ is the effective void radius, $[3/(4\pi) V]^{1/3}$.  Also, we explicitly remove voids with $\rho_{\rm min} \ge 1$ \red{(where $\rho_{\rm min}$ is the \red{minimum} density of a void, and a density minimum is a galaxy with lower density than any of its neighbours)}, since for our particular application, we do not want to use local density minima within obviously overdense regions. 

To deal with the boundaries, we use the same set of buffer galaxies surrounding the survey as used by \citep{Sutter12} (in their November 2012 catalog).  \citet{Nadathur2013} have found that this set of buffer galaxies is sparser than would be ideal, and also that the positions of some bright stars were not considered in the analysis. These could result in some galaxy densities being corrupted. However, if these corrupted densities result in spurious voids, they \red{should typically be eliminated by our strict cuts} in void effective radius and minimum-to-ridge density contrast, as we describe below.  \red{\citet{Nadathur2013} have also raised the concern that void detections can be unexpectedly sensitive to changes in the distance coordinate. Ideally, perhaps, we would have used a $\Lambda$CDM distance estimate instead of the redshift distance $cz$. However, again, we do not expect the choice of distance coordinate to matter substantially for the large voids that survive our conservative cuts.} Also, importantly, any corruptions should only contribute to the noise, rather than to the signal, in our measurement.

\section{Simulations}
\label{simulation}

\begin{figure*}
\begin{center}
    \advance\leftskip 0.40cm
    \scalebox{0.43}{
     \includegraphics[angle=0]{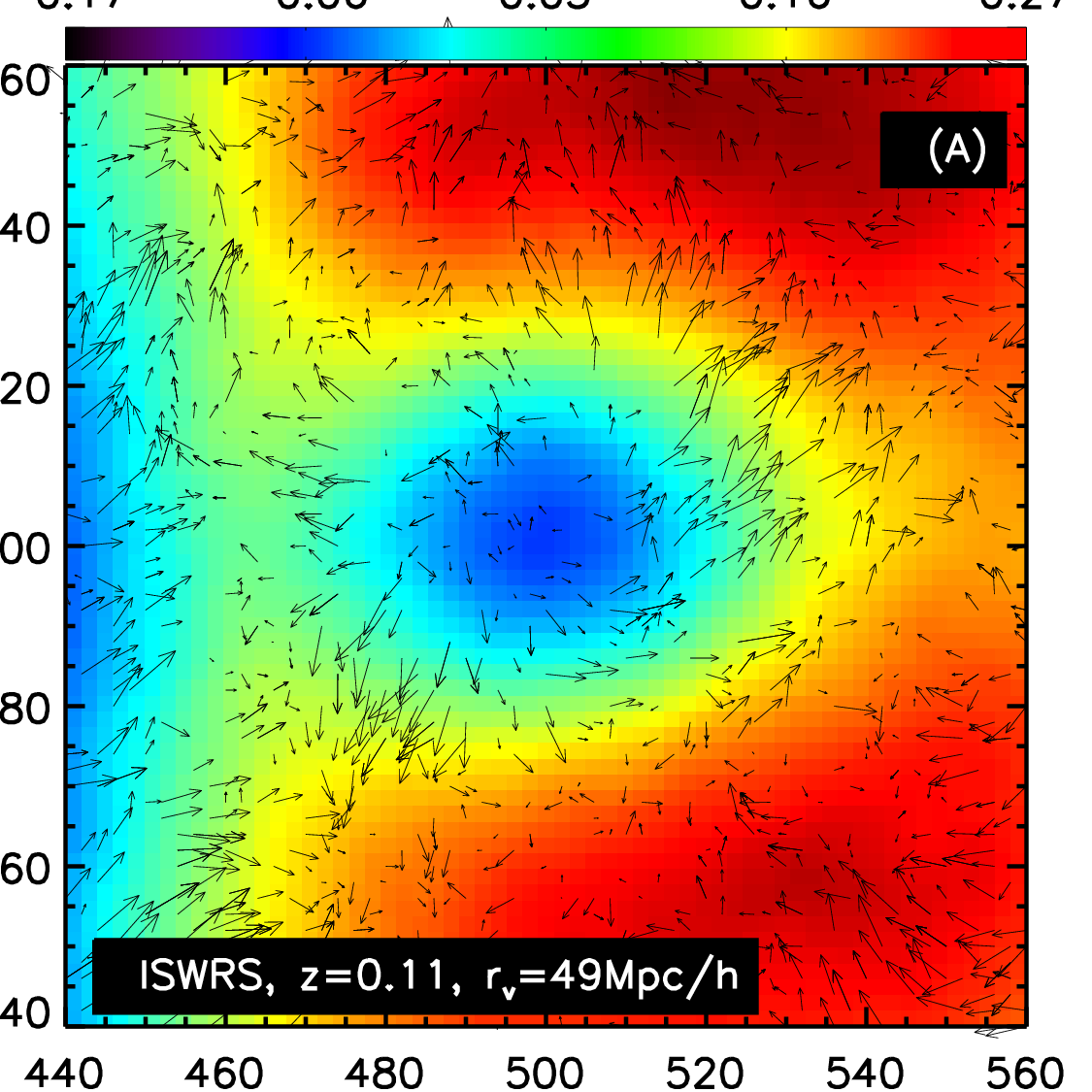}}
\hspace{0.6 cm}
   \scalebox{0.43}{
      \includegraphics[angle=0]{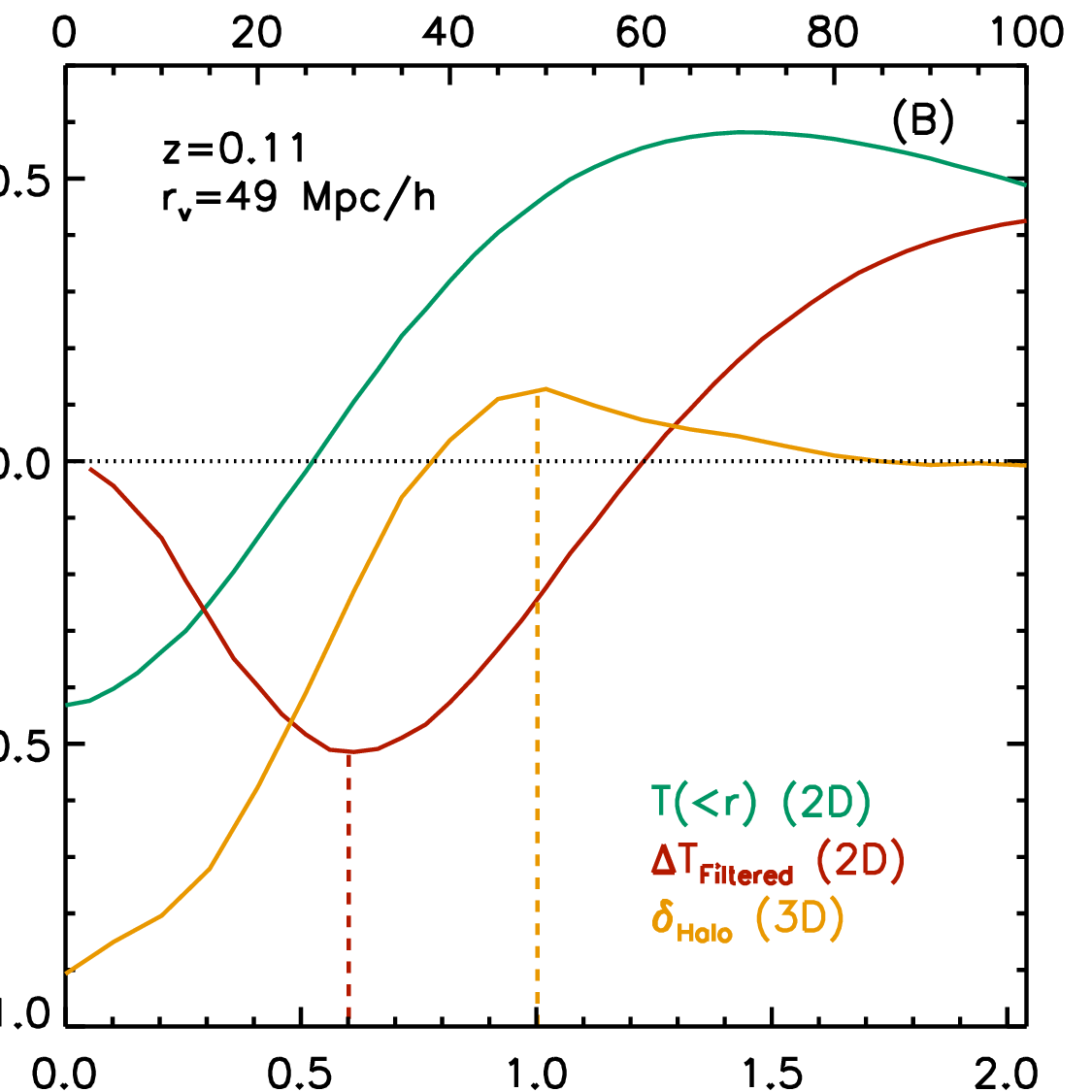}}
\vspace{1.0cm}
    \caption[]{From $N$-body simulations, the stacked ISWRS 2D temperature (A) and the 1D average cumulative temperature profile (green-solid on panel-B) 
for simulated voids with radius $r_{\rm v}\approx 49$ Mpc/$h$. Arrows are the mass-weighted velocity of the dark matter, indicating 
the outward expansion of matter from the centre of the ISW cold spot. In panel-B, 3D void 
density profiles traced by halos are shown in orange. The red-solid line is 
obtained from convolving the average 2D projected ISWRS temperature map with compensated top-hat filters of different radii $R$. The red-dash line indicates the filter 
radius where the ISWRS temperature reaches its minimum (maximum in absolute value). It is about $60\%$ of the effective void radius traced by halo number density, as indicated 
by the orange-dash line. The zero-crossing of the green curve is very close 
      to the radius where the filtered ISWRS signal peaks. These profiles come from the stacking of 475 voids within the 
      radius of 45 to 55 Mpc/$h$ identified using halos at $a=0.9$ from the simulation with the box size of L=1000 Mpc/$h$. For the ISWRS maps, Fourier modes with 
$k<0.01h$/Mpc are removed to reduce the noise. The temperatures on the B panel are multiplied by a factor of 5 for better illustration.
    }
    \label{Tsimulation}
\end{center}
  \end{figure*}

For testing our analysis pipeline and to understand the expected ISWRS signal, we construct mock $\Lambda$CDM void catalogs 
from $N$-body simulations and compute their expected ISWRS signal. Our goal is to perform the 
same analysis in the SDSS data as in our simulated voids, where the ISWRS signal is known. 
\subsection{Simulations of voids} 
We construct mock void catalogs using a set 
of simulations run in the concordance cosmology ($\Omega_{\rm m}=0.24, \Omega_{\Lambda}=0.76, n_s=0.958, \sigma_8=0.77, h=0.73$) \citep{Li2013}.  
The simulations are run with the following box sizes and numbers of particles: [$L$=1500 Mpc/$h$, $N_{\rm p}$=1024$^3$), 
[$L$=1000 Mpc/$h$, $N_{\rm p}$=1024$^3$) and [$L$=250 Mpc/$h$, $N_{\rm p}$=512$^3$]. We use halos matched to the number 
densities of galaxies in the 6 volume-limited sub-samples of the SDSS data. We use halos with more than 20 particles, as linked by the Friends-of-Friends 
algorithm with the linking length of 0.2 \citep{Davis1985}. The particle masses in our simulations are $2.09\times 10^{11} {\rm M_{\odot}}/h$, $6.20 \times 10^{10} {\rm M_{\odot}}/h$, and $0.77\times 10^{10} {\rm M_{\odot}}/h$, giving 
minimum halo masses $\rm {M_{\rm min}}= 4.18\times 10^{12}{\rm M_{\odot}}$/$h$,  $1.24 \times 10^{12}{\rm M_{\odot}}/h$, and $1.54 \times 10^{11}{\rm M_{\odot}}/h$.  Halos are approximated 
as galaxies, assuming that each main halo hosts one SDSS galaxy. 
This simple treatment neglects the complexity of galaxy formation and halo occupation.  In the densest galaxy samples, for example, large halos should host multiple galaxies, better delineating void edges.  Excluding these extra galaxies may reduce our ability to detect voids in the simulations.  But
given that we rely on halos just to find voids, and that the typical sizes of voids are usually orders of magnitude larger than the size of halos, the simulated void catalog should be acceptable for our purposes.  

To model the signal, it is important to match the galaxy sampling density, since more, and smaller, structures are found with increasing sampling. 
We adjust $M_{\rm min}$ to match the number density of galaxies except in the two lowest-redshift
sub-samples, where the number density of galaxies are beyond the
resolution limit of our current simulations. In principle, we could use a
higher-resolution simulation to match these densities. But given that the volume of these two sub-samples
is less than $2\%$ of the total, we expect them to contribute very
little to the final stacked signal, which will be demonstrated in the
next section.  We therefore do not make the effort to analyze higher
resolution simulations, and leave it for future work. The two 
highest-redshift LRG subsamples have the largest average 
voids and dominate the volume. To reduce cosmic variance in the average signals, 
we employ 6 realizations of the simulation with $L=1500$ Mpc/$h$. 
 
We apply the same void finding algorithm as in the real data to these mock halo catalogs 
and find voids at 4 discrete redshift slices, $a=${0.7, 0.8, 0.9, 1.0} in the cubic simulation boxes, 
where $a=1/(1+z)$ is the expansion factor. This covers the whole redshift range of the SDSS void catalog. 

\subsection{Simulations of ISW}

For each simulation box, we follow the algorithm presented by \citet{Cai10} [see also \citep{Seljak96,Smith2009, Nadathur2012, HM12}] to compute the time derivative of the potential $\dot\Phi$ 
using particle data. This can be achieved by computing $\dot\Phi$ in Fourier space using
\begin{equation}\label{eq4}
\dot{\Phi}(\vec{k},t)=\frac{3}{2}\left(\frac{H_0}{k}\right)^2\Omega_{\rm m}
\left[\frac{\dot{a}}{a^2}\delta(\vec{k},t)+\frac{i\vec{k}\cdot\vec{p}(\vec{k},t)}{a}\right],
\end{equation}
where $\vec p(\vec k,t)$ is the Fourier transform of the momentum density divided by the mean mass density, 
$\vec p(\vec x,t)=[1+\delta(\vec x,t)]\vec v(\vec x,t)$, and  $\delta(\vec k, t)$ is the Fourier transform of the density contrast. 
$H_0$ and $\Omega_{\rm m}$ are the present values of the Hubble and matter density parameters. The inverse Fourier transform of the above 
yields $\dot\Phi$ in real space on 3D grids. The integration of $\dot\Phi$ along the line of sight yields the full ISWRS temperature fluctuation
\begin{equation}
 \frac{\Delta T(\hat n)}{T}=\frac{2}{c^2} \int \dot \Phi(\hat n, t) dt. 
 \end{equation}
   
In the linear regime, the velocity field is related to the density field by the linearized continuity equation 
$\vec{p}(\vec{k},t)=i{\dot\delta(k,t)\vec k}/{k^2}\approx i{\beta(t)\dot a/a\delta(k,t)\vec k}/{k^2}$. Thus
\begin{equation}\label{eq5}
\dot{\Phi}(\vec{k},t)=\frac{3}{2}\left(\frac{H_0}{k}\right)^2\Omega_{\rm m}
\frac{\dot{a}}{a^2}\delta(\vec k,t)[1-\beta(t)],
\end{equation}
where $\beta(t)$ denotes the linear growth rate
$\beta(t)\equiv{d \ln D(t)}/{d \ln a}$.
This equation models the linear ISW effect, using only information from the density field.
This approach has also been adopted by \cite{Watson2013} for larger volume simulations.

\section{Stacking of Voids}
\subsection{Stacking in simulations}
With the simulated void catalogs and the simulated ISWRS and ISW maps, we can do the same stacking as in the observations, 
and make predictions of the expected ISW signal in the concordance cosmology. 

We follow a similar procedure as G08 for stacking superstructures, with the key difference being that we scale the filter radius to each void.  In G08, a constant filter radius was used because photometric redshift errors prevented accurate estimates of physically meaningful void radii. In contrast, the current sample allows more
accurate estimates.

The filtered CMB signal around each void is the temperature averaged over a circular aperture $r<R$ around the void centre, minus the temperature averaged over a surrounding annular aperture $R<r<\sqrt{2}R$:
\begin{equation}
\Delta T = \bar T_1-\bar T_2 = \frac{\int_0^R T(\vec r) d\vec r}{\int_0^R d\vec r} - \frac{\int_R^{\sqrt{2}R} T(\vec r)d\vec r}{\int_R^{\sqrt{2}R} d \vec r},
\end{equation}
where $R$ is the radius of the filter. We use $\Delta T$ for filter temperature and reserve $T$ for the unfiltered temperature throughout the paper.
The main purpose of this filter is to suppress large-scale power contamination from the primordial CMB.
In reality, void profiles are not compensated top-hats, and this filter is likely not the optimal for ISW detection.  However, as long as we perform the same convolution for both the real data and simulations, direct comparisons between them are fair and meaningful. This relatively simple convolution 
enables us to easily adjust the size of the filter according to the size of each void; the optimal ratio of the filter size $R$ and the \zobov\ void size is determined from simulations. 

With this aim, simulated voids of similar radii at the same redshifts are stacked. This is done for both the halo number-density 
field and the 3D ISWRS signal, i.e. $\dot\Phi$ on a 3D grid. Then, the stacked 3D grids of $\dot\Phi$ are projected along all three axes of the cubic simulation box. 
This yields the 2D stacked ISWRS temperature $\Delta T$ map, as shown on Fig.~1-A. A cold spot corresponding to the stack of 475 voids is clearly seen, as expected.
The outflow of dark matter indicated by the mass-weighted velocity field (shown in arrows in Fig.~1-A) explicitly shows that the stacked void region is expanding.
The map is then convolved with compensated top-hat filters of different radii, from which filtered ISWRS temperatures corresponding to the void region are found. 

Here we see that quasilinear or nonlinear-scale voids generally have overdense shells around them. Indeed, that is roughly the definition of a \zobov\ void. 
The ISWRS signals of these nonlinear voids generally have hot rings around them, as shown in the green curve of Fig.\ \ref{Tsimulation}.  So, although it 
has been argued that an uncompensated filter is of equal value in detecting features in the CMB \citep{Zhang2010}, at least for detection of an ISW-like void or 
supercluster imprint, use of a compensated filter is justified, and greatly preferred. 

\subsection{Optimizing the filter size}
\label{simulation}

\begin{figure*}
\begin{center}
    \advance\leftskip 0.40cm
    \scalebox{0.43}{
     \includegraphics[angle=0]{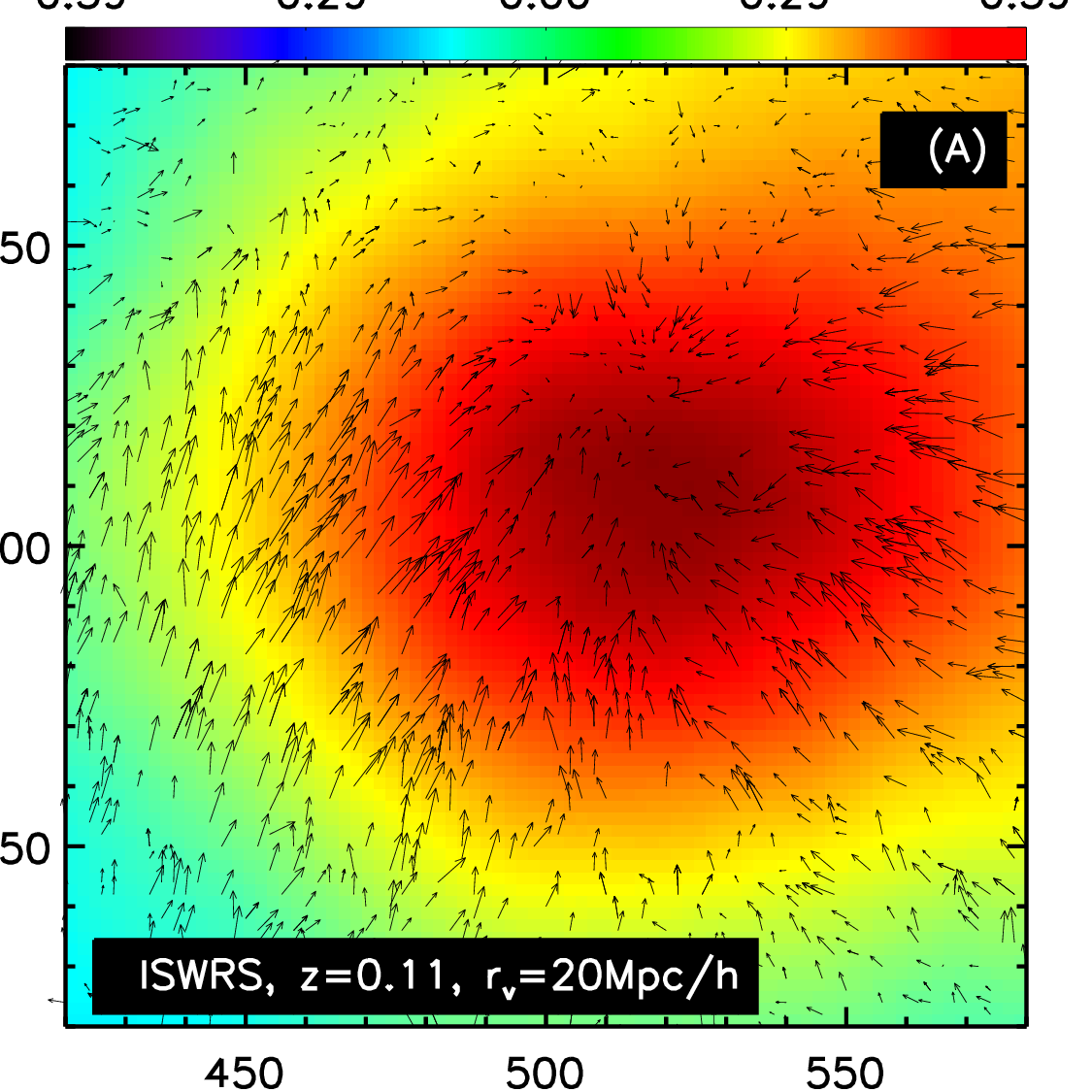}}
\hspace{0.6 cm}
   \scalebox{0.43}{
      \includegraphics[angle=0]{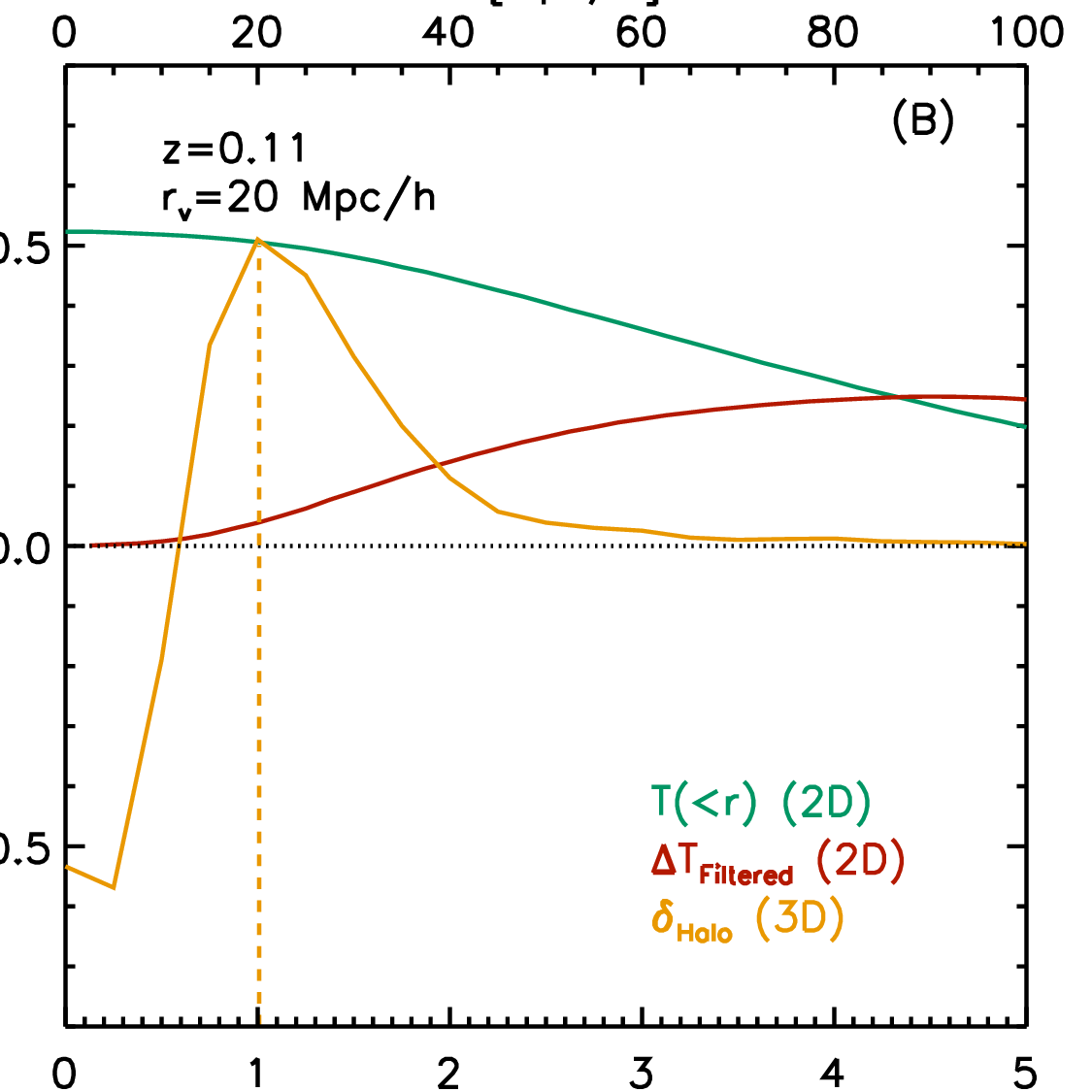}}
\vspace{1.0cm}
    \caption[]{The same as Fig.\ \ref{Tsimulation}, but stacking simulated voids with radius $r_{\rm v}\approx 20$ Mpc/$h$.
There are 
626 voids with radius between 17.5 and 22.5 Mpc/$h$; they are identified using halos at $a=0.9$ from 
the $L$=1000 Mpc/$h$ simulations. Even though they are true underdensities in the halo number density, as shown in the orange line of panel B, 
their stacked ISWRS signal gives a hot spot, as shown in panel A, and also in green line of panel B. Arrows in panel-A are the corresponding 
mass-weighted velocity of the dark matter, indicating the convergent flow of matter towards the centre of the local void region defined by halos.}
    \label{Tsimulation2}
\end{center}
  \end{figure*}
To find the optimal filter radius for a given void effective radius, we explore a wide range of filter 
radii in our simulations. Fig.\ 1-B shows the filtered ISWRS temperature versus filter radius $R$ for voids of the average radius $r_{\rm v} \sim 49$ Mpc$/h$ (red line). 
We also plot the cumulative ISWRS temperature profile without convolution with the filter in green and the 3D halo density profile in orange. Interestingly, we find that the filtered ISWRS 
signal peaks at $R \sim 0.6 r_{\rm v} $, i.e.\ at significantly smaller values than the measured void radius, roughly coinciding with the zero-crossing of the ISWRS temperature profile. 
The optimal filter size appears to be independent of redshift. It is a weak function of the void radius,
increasing to approximately $0.7$ for $r_{\rm v}>70$ Mpc$/h$. For the SDSS analysis, we implement
this nearly universal filter size as a ``rescaling factor'' before stacking.
 
Compared to the density field, the potential (and its time derivative) carries an extra factor of $1/k^2$, causing scales much larger than the voids in the catalog to dominate the ISWRS temperature maps. To reduce this large-scale variance, we remove some very large-scale modes ($k<0.01h$/Mpc) when showing the ISWRS temperature map and profiles.
Even with the $k$-mode removal and the stacking of 475 voids, the ISW cold spot still does not seem very circular; there is still substantial noise.  Indeed, when analyzing subsamples, we found many fluctuations in the 2D map (Fig.\ 1-A) and in the uncompensated average temperature as a function of radius (the green line of Fig.\ 1-B). However, the compensated-filtered temperature in Fig.\ 1-B remained quite stable; this is another justification for using a compensated filter. It is reassuring that the halo-density profile (yellow line) matches expectation for a void profile.

\subsection{Optimizing the void catalogue}
Naively, to avoid $a$ $posteriori$ bias, one may simply take the entire void catalog for stacking the CMB. 
However, this is risky in that A.) voids found in galaxy catalogs might not correspond to real voids in the density field. 
If the sampling is sparse, only large voids can be detected. 
B.) There are voids whose sizes are about the same as the mean galaxy separations for 
each volume limited sample, which may be spurious.
C.) It is known that some voids, in particular small ones, may live in over-dense environments, where the large-scale 
environment might be contracting. These so called `voids-in-clouds' \citep[e.g.][]{ShethVDWeygaert2004, Ceccarelli2013, HamausEtal2013} could generate ISW hot spots on a 
larger scale than the detected void, and it is unclear that our compensated filtering would be sufficient to pick out a cold spot.

We note also a previously unappreciated reason why the smallest voids in a halo sample might tend to reside in larger-scale overdensities. 
Large-scale overdense patches have higher halo number density, and thus higher sampling than average. High sampling is exactly what is needed to resolve the 
smallest voids, so it is not surprising that a sample of only the smallest voids would tend to be in large-scale overdense regions.

Fig.\ \ref{Tsimulation2} shows an ISWRS stack using relatively small voids, $r_v\sim20$Mpc/$h$. We find from the stacked halo number density 
profile (Panel B) that the voids are indeed real underdense regions. However, stacking their CMB imprints yields a 
large hot spot with the radius of more than 80 Mpc/$h$. This suggests that those relatively small voids are likely to live in overdense environments 
which are contracting. Indeed, the mass-weighted velocity field overplotted in Panel A indicates convergent flow of 
dark matter towards the stacked void centre. \red{(We notice that the large-scale hot spot is slightly off-centred, most likely due to cosmic variance introduced by large-scale modes.)} 
It is not ideal to blindly take voids defined in the halo density field for the ISWRS 
stacking procedure. The void-in-cloud problem may reduce the total stacked signal, or even reverse the sign of it. It also complicates the interpretation 
of the stacked ISWRS signal. To tackle this problem, we turn to our simulations.

In simulations, we stack voids in relatively narrow ranges of radii 
(5 Mpc/$h$) at each redshift of our simulations, applying the optimal rescaling filter radius for the compensated filter to obtain the 
stacked, filtered ISWRS temperature. Results are shown in Fig.\ \ref{Trz}. The ISWRS temperature clearly depends on size and redshift, 
i.e.\ larger voids induce a greater ISWRS signal. A void at constant $r_{\rm v}$ has larger $|\Delta T|$ at larger scale factor, indicating 
the increasing influence of dark energy.  Voids with $r_{\rm v}<20$ Mpc/$h$ have $|\Delta T|<0.1\mu K$ at all redshifts. The largest 
voids found in the SDSS data ($r_{\rm v}\sim 140$ Mpc/$h$) may have $|\Delta T |\sim 1-2\mu K$, but they are very rare. The magnitude of the 
linear ISW signal is typically 10$\%$ to 20$\%$ lower than the full ISWRS signal. Larger differences are found for smaller voids such as 
$r_{\rm v}<20$ Mpc$/h$, but the overall amplitudes there are negligibly small. 

Notice that each colored curve crosses zero at low void radius, and stays close to zero. 
This is an indication of the effect shown in Fig.\ \ref{Tsimulation2}, or could possibly be a sign of spurious voids. To get rid of them, we draw cuts in void radius based on the zero-crossings of those simulated curves 
for each volume-limited sample. They are $r_{\rm cut, default}=$[20, 25, 30, 35, 45, 65] Mpc/$h$, where two of them are from interpolations. 
Unfortunately, the majority of the voids in DR7 are smaller than those sizes. With these cuts, we throw away 2/3 of 
the 1521 voids, retaining only 477 voids. If the signal corresponds to what we find in simulations, this should ensure that on average, the stacked ISWRS signal for voids is negative.

These cuts may seem over-conservative, but the expected S/N for 
the ISWRS signal from an individual void is so small that it can be easily swamped by the noise. It is worth making an aggressive cut to 
reduce the noise if we have good physical reasons. We will investigate this issue further 
in section \ref{tests}. In the next subsection, we apply these cuts to the real data and present our main results on these relatively clean void catalogues.

\subsection{Stacking with SDSS data} 
In individual voids, the expected ISWRS signal we are interested in is overwhelmed by two major sources of noise, 
1.) the primordial CMB temperature fluctuations, and 2.) the ISWRS temperature fluctuations that have larger coherent scales 
than the typical size of our voids. These noise sources, which are essentially cosmic variance, are much greater than the 
ISWRS signal for which we are looking. It is therefore necessary to try to suppress them to see the potential ISWRS signal. 
For this purpose, we use two techniques. First, we remove large-scale modes from the CMB, 
i.e. $\ell \leq 10$, knowing that these scales are much larger than the sizes of our voids and the sought ISWRS signal will not 
be affected by the removal of them. All results we show in the rest of the paper have the $\ell \leq 10$ restriction unless 
specified otherwise. Second, we apply compensated top-hat filters to the CMB, hoping to further reduce the influence of 
large-scale modes. Of course, stacking a large number of voids can also help to reduce the noise, but we are limited by the size of the 
current data.

We use the WMAP9 foreground-reduced 
Q, V and W frequency maps \citep{WMAP9} for the stack, excluding voids which overlap
by $>20\%$ with WMAP masked regions. This reduces the number of voids slightly, from 477 to 470. We then use the void centers for stacking the CMB 
maps, rescaling the CMB according to the effective radius of each void, and applying the compensated filter to it. The filter for each void is scaled 
to $60\%$ of the radius, the scaling factor found in simulations. The solid lines on the top panel of Fig.\ \ref{detection} show the cumulative stacked CMB 
temperature $\Delta T$ versus the number of stacks. The stacked CMB
temperature is negative for almost all $n_{\rm stack}$, an indication of the stability of the signal. The stacked temperature varies from -2$\mu$K to
-3$\mu$K after stacking $\sim$150 voids. When all the 470 `cleaned' SDSS voids are used for stacking, the resulting filtered temperatures show little frequency dependence; they are 
{-2.6}, {-2.8} and {-2.8} $\mu$K in the Q, V and W bands, respectively. This is consistent with an ISWRS-like signal.
\begin{center}
\begin{figure}
\advance\leftskip -1.2cm
\scalebox{0.51}{
\includegraphics[angle=0]{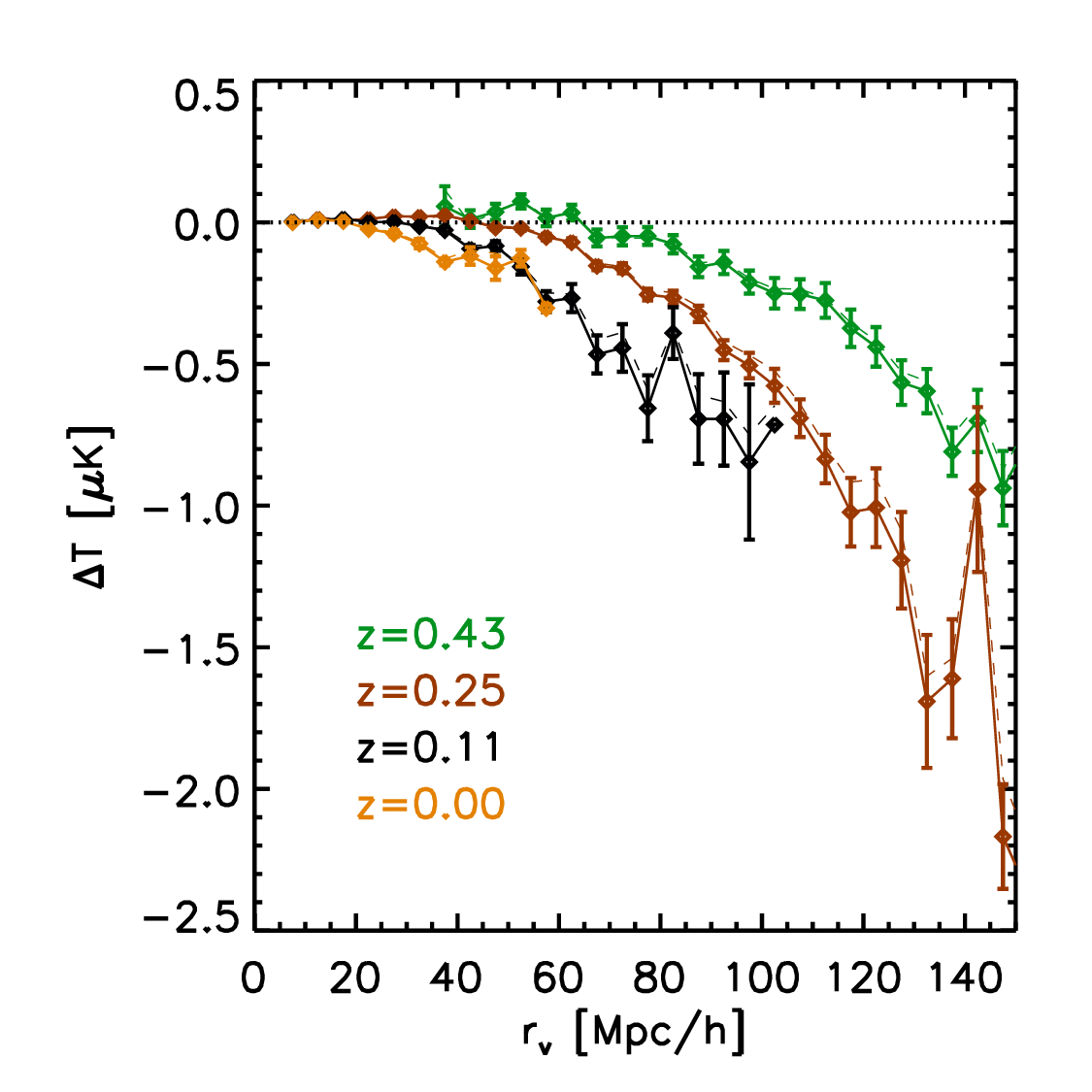}}
\caption{Filtered ISWRS temperature, $\Delta T$, for voids of different radii, $r_v$, at different redshifts as indicated in the legend. These results are from the 
simulations but measured as in the observations. Dashed lines indicate the linear ISW signal. 
We linearly interpolate the zero-crossings of these curves to determine the cuts in void radius to apply to the SDSS DR7 void catalog.}
\label{Trz}
\end{figure}
\end{center}

\begin{figure}
\vspace{.7 cm}
\advance\leftskip -1.3cm
\scalebox{0.5}{
\includegraphics[angle=0]{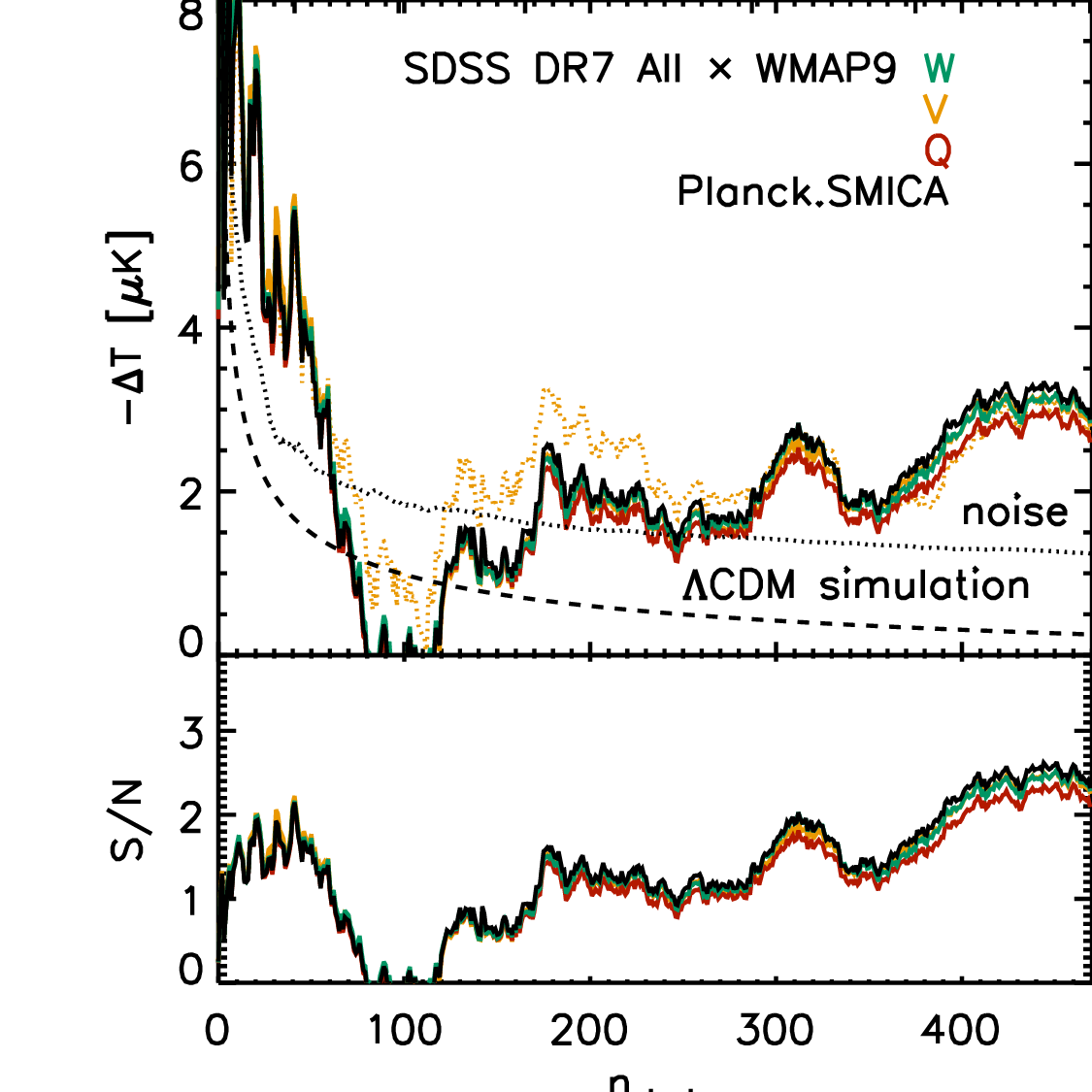}}
\vspace{.7 cm}
\caption{Stacking of WMAP9 Q(red),V(orange),W(green)-band and Planck SMICA maps using void catalogs from the SDSS DR7 galaxy sample. In the upper panel, solid curves show cumulative stacks of the compensated-filtered CMB temperature. Voids are sorted by size, which increases from 
the left to the right. \redd{The orange dotted curve is similar to the orange solid curve, but each void is weighted by its 
probability of not being random in a Poisson model, according to Eq.\ref{eqn:poisson}. See Fig.~\ref{1Dparameter} and section~\ref{tests} for more detail.} The black-dotted line is the variance calculated from randomizing the sky position of each void 1000 times and repeat the stacking.  The dashed line is the stacked linear+nonlinear ISWRS signal in the fiducial concordance cosmology using simulated void catalogs 
described in section \ref{simulation}.  The lower panel shows the cumulative signal-to-noise ratio (S/N) for each stack. After stacking about 200 voids, 
we get a $\sim 2 \sigma$ detection.}
\label{detection}
\end{figure}
To estimate the statistical significance, we measure the noise in two ways. First, we keep the sizes and relative positions of all voids fixed and randomly rotate them on 
the CMB map before stacking. The WMAP mask is applied in the same way as we do for the original stacking. We estimate 
the variance of the stacked temperatures from 5000 such randomizations. The resulting 1$\sigma$ variantion is shown as the dotted curve in 
Fig.\ \ref{detection}. Second, we use the best-fit CMB power spectrum \citep{Larson11} to generate 5000 mock CMB maps, and repeat the same stacking procedure 
with them. The variances of the random sample are nearly the same as those from the first method, at the percent level. This indicates the robustness of 
noise estimation. 

The signal-to-noise ratio (S/N) of the stack is shown in the bottom panel of Fig.\ \ref{detection}. With a certain amount of random fluctuation, the S/N 
is in general increasing with the number of voids in the stacks. After stacking $\sim$150 voids, it reaches $\sim$2$\sigma$. 
The stack of all cleaned SDSS voids yields $\sim$$2.1$, 2.2, 2.2$\sigma$ for WMAP9-Q, V and W bands. 

We have also tried using the Planck SMICA map \citep{PlanckXII}, with the same WMAP9 mask, for the stacking, and find $\Delta T\sim -2.9 \mu$K, which is 
about 2.3$\sigma$, slightly higher than WMAP results, but the differences are negligible. The black-solid lines in 
Fig.\ \ref{detection} shows that the cumulative stacking signal with Planck is also very similar those from the WMAP bands. 
It is reassuring that switching from WMAP9 maps to Planck SMICA, with much higher resolution ($N_{\rm side}=2048$) yields the same results.

Qualitatively, this negative temperature fluctuation corresponding to the stacking of voids is 
expected from the linear ISW effect in the $\Lambda$CDM Universe. Therefore, this signal, although it is marginal in significance, 
suggests that the Universe at $0<z<0.44$ is accelerating in its expansion. We make quantitative comparison of the signal 
with that expected in $\Lambda$CDM using our simulations. From the simulated mean filtered ISWRS temperature $\Delta T$ for voids of 
different sizes at different redshifts presented in Fig.\ \ref{Trz}, we sample from the table of $\Delta T(r_{\rm v}, z)$ for 
the 470 voids with the same sizes and redshifts as in the SDSS data, and obtain the simulated curve of cumulative stacked $\Delta T $ 
versus the number of voids in the stack for this cosmology shown by the dashed line in Fig.\ \ref{detection}. The magnitude of the 
simulated ISWRS signal is substantially smaller than the observed one. If there is no other contamination or systematics, the data 
suggest a $\sim 2 \sigma$ tension with the $\Lambda$CDM cosmology. 
\begin{figure}
\advance\leftskip -0.5cm
\scalebox{0.5}{
\includegraphics[angle=0]{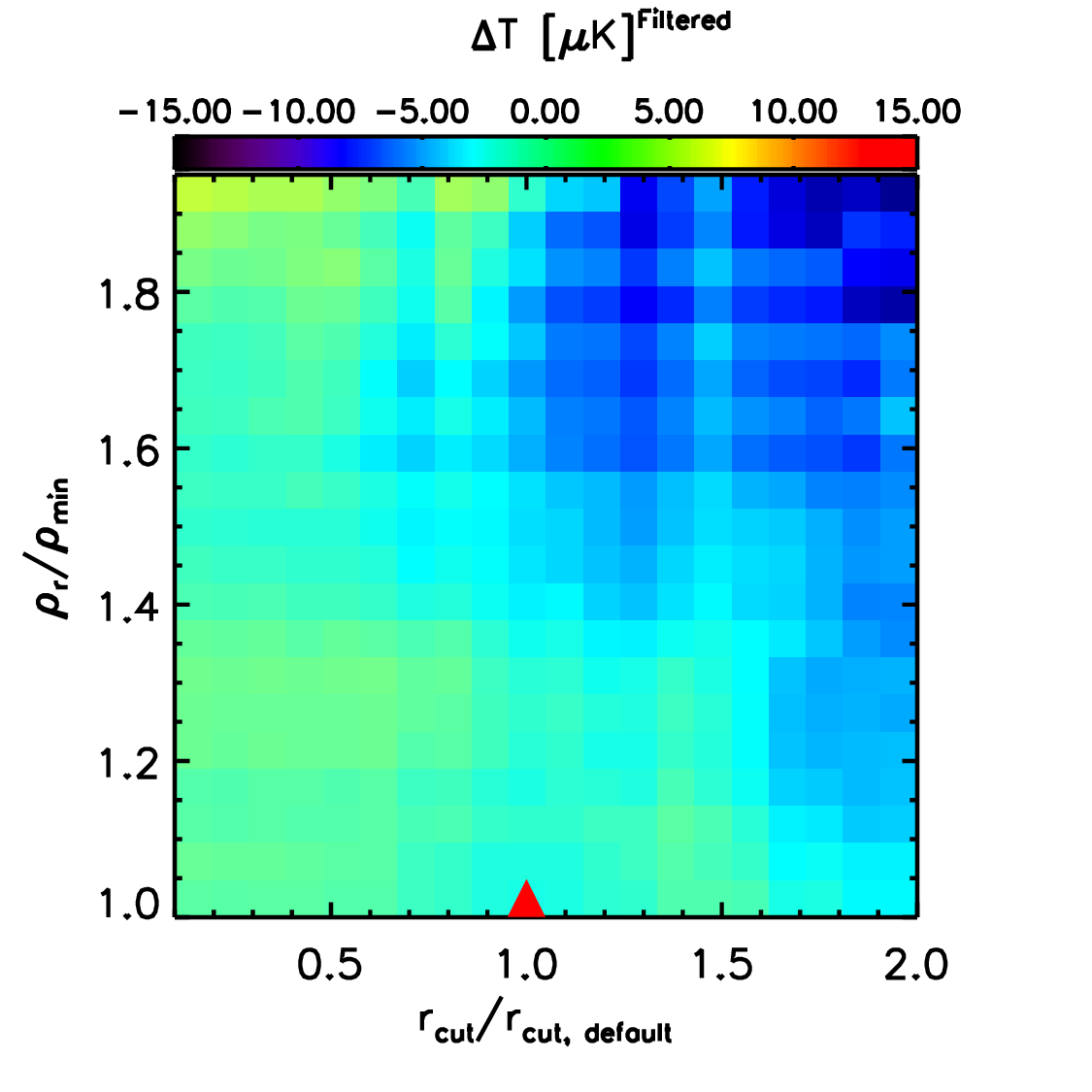}}
\caption{Filtered-stacked temperature as we change the cuts for the void sizes and probability of voids around the fiducial 
values chosen from simulations. The $x-$axis is the threshold for void sizes compared to the fiducial cuts;  on the right, only the largest voids are kept. The $y-$axis is a threshold in a measure of the statistical significance of a void. Higher $\rho_{\rm r}/\rho_{\rm min}$ is more significant. \red{The red triangle indicates the result from the default cuts of $r_{\rm cut}/r_{\rm cut, default}=1$ and $\rho_{\rm r}/\rho_{\rm min}=1$. 
Its value corresponds to the value of $\Delta T$ where the yellow curve ends at the right of Fig.\ \ref{detection}. }}
\label{2Dparameter}
\end{figure}

\begin{figure*}
\begin{center}
\advance\leftskip -0.6cm
    \scalebox{0.34}{
     \includegraphics[angle=0]{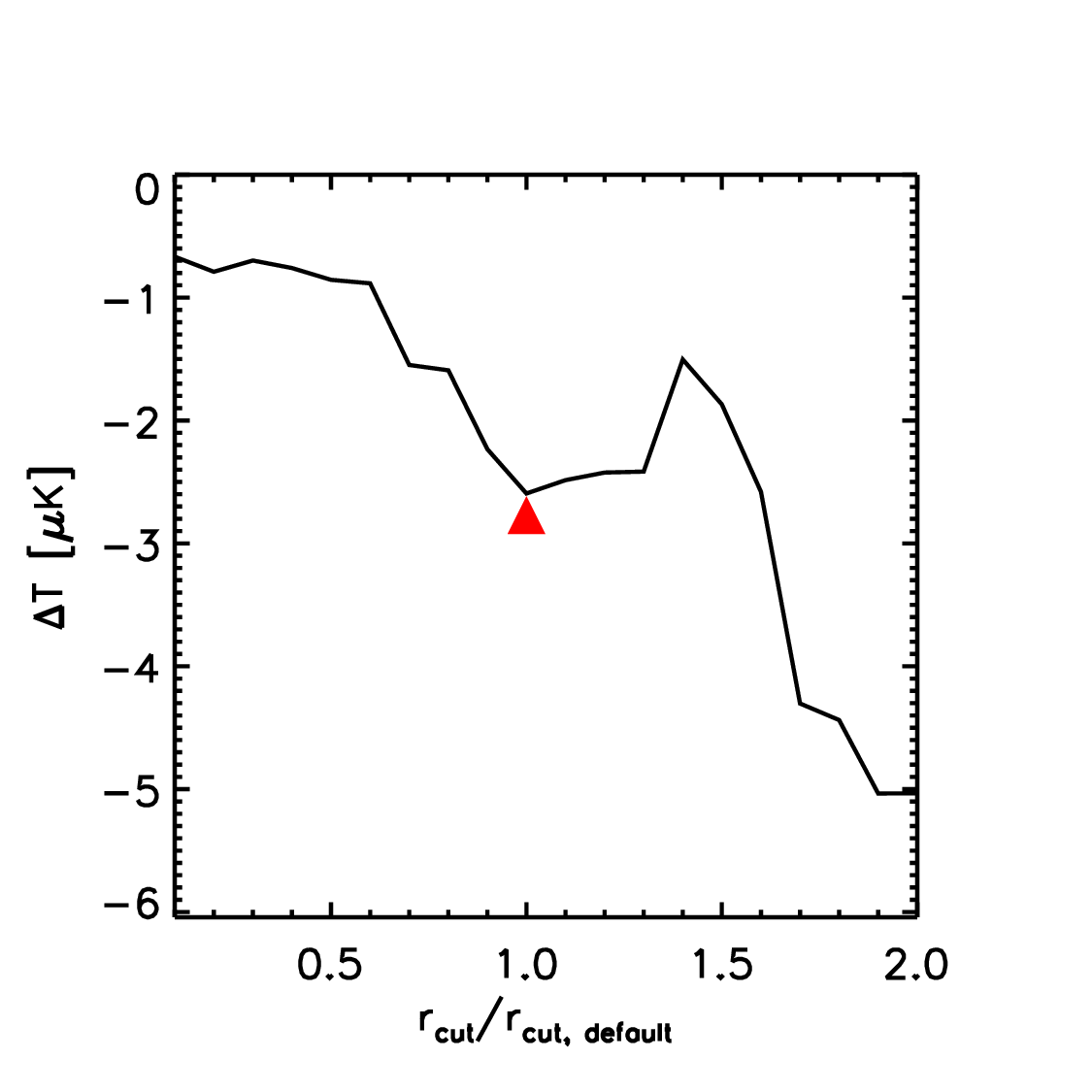}
      \includegraphics[angle=0]{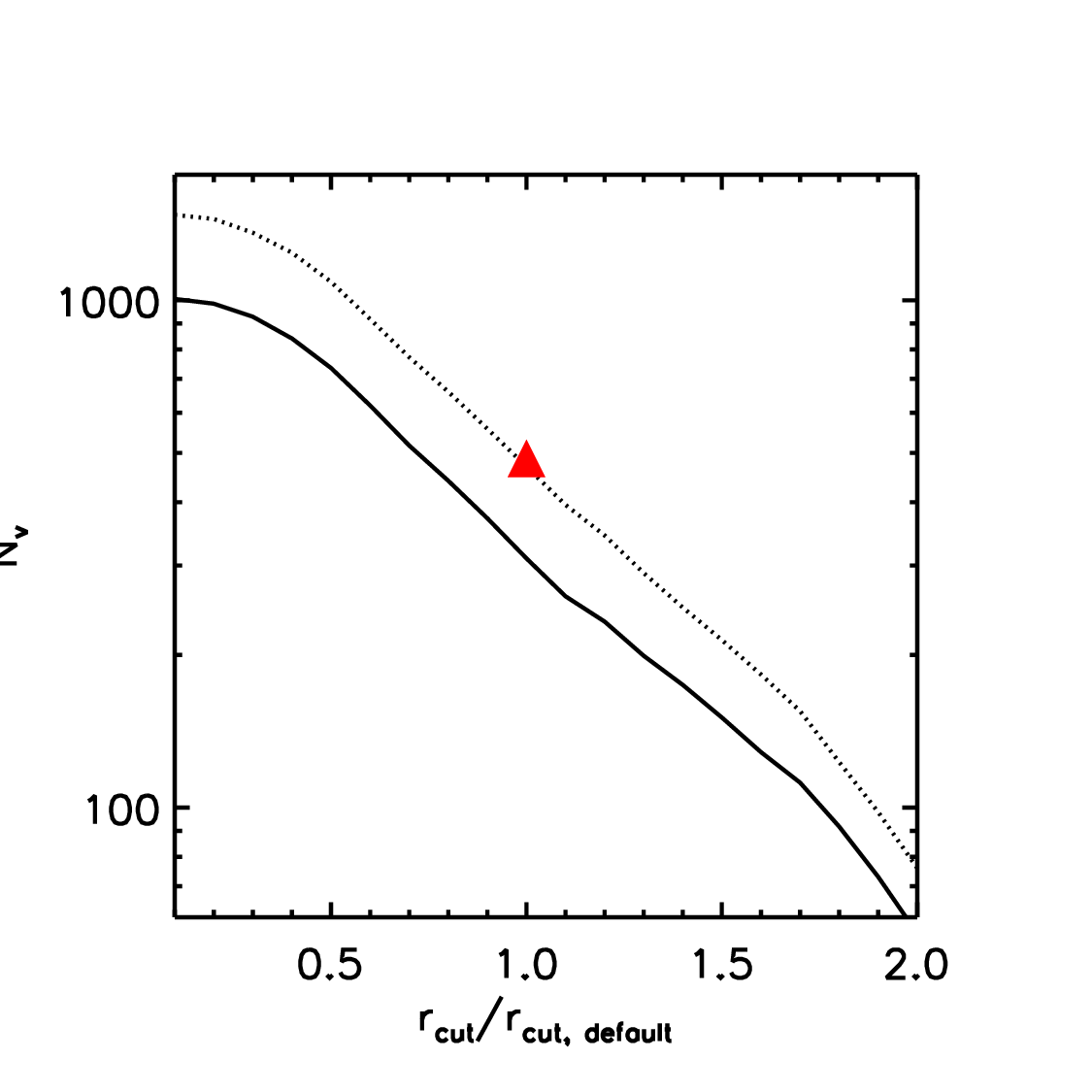}
      \includegraphics[angle=0]{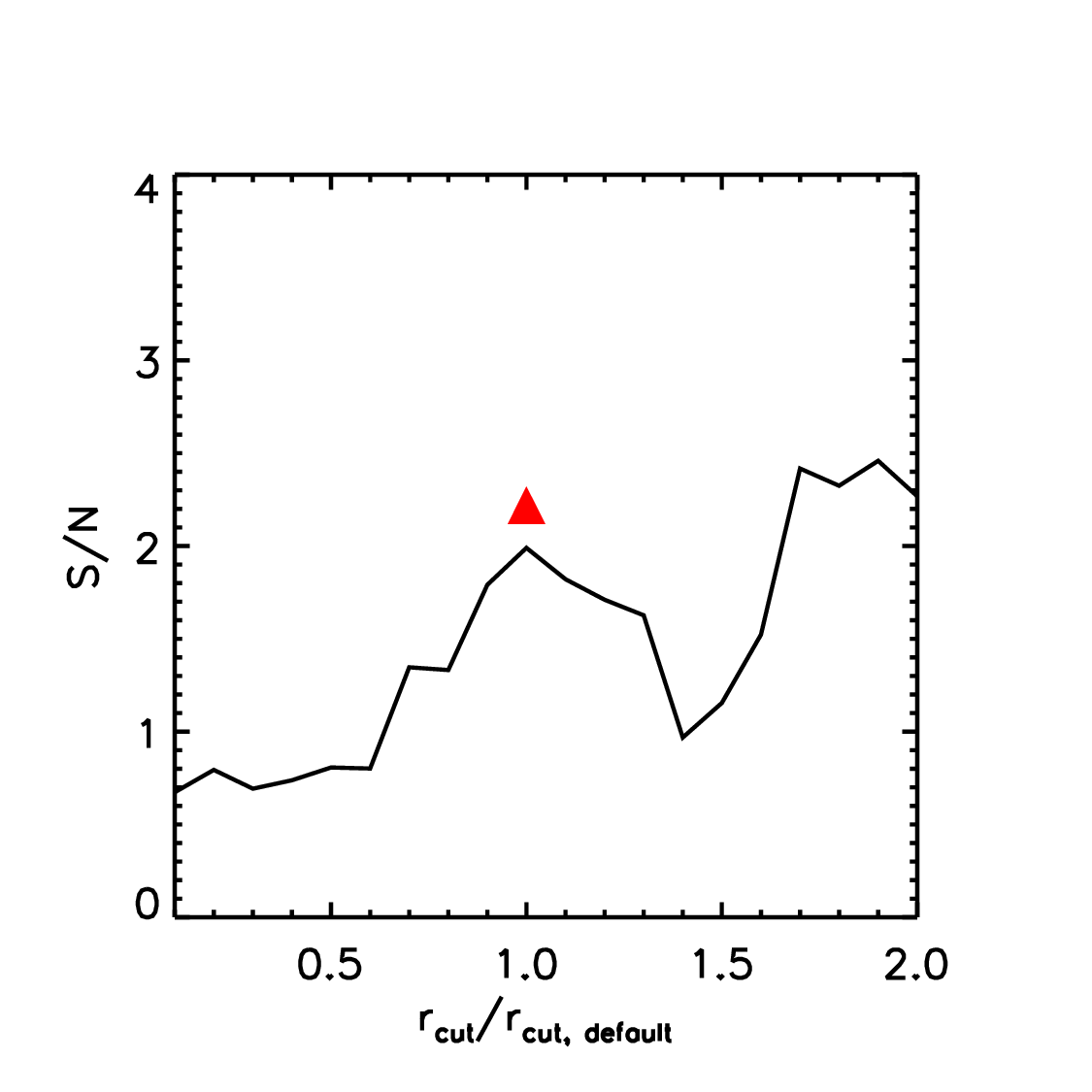}}
\caption{Left: Filtered temperature versus minimal void radius, each void is weighted by its probability of not being random in a Poisson model. 
Middle: number of voids remain in the sample as minimal void radius increase (solid), and the effective number of voids (void-probability weighted) 
(dotted). Right: estimated Signal-to-Noise ratio. \red{The red triangle in each plot indicates the default cuts of $r_{\rm cut}/r_{\rm cut, default}=1$ and $\rho_{\rm r}/\rho_{\rm min}=1$. Its value on the left-hand panel 
correspond to value of $\Delta T$ where the yellow curve ends at the right of Fig.\ \ref{detection}. Its value at the middle figure is the number of voids left after applying the default cuts.
Its value on the right-hand panel corresponds to the final S/N in Fig.\ \ref{detection}. } }
\label{1Dparameter}
\end{center}
  \end{figure*}
  
We caution that our $\sim2\sigma$ measurement of cold imprints of voids on the CMB is difficult to explain with the ISW effect in a $\Lambda$CDM universe. 
Suppose in an optimistic case, the expected ISWRS signal is of the order of $1~\mu$K for voids of $r_{\rm v} \sim 100$ Mpc/$h$, as shown in 
Fig.\ \ref{Trz}, and the CMB temperature fluctuation is of the order of $30 ~\mu$K. Assuming the noise in the 
stack goes down as $1/\sqrt{N}$, one will need a stack of 8100 voids to 
make a $3\sigma$ detection. This is beyond the reach of any existing data. As another example, if we have voids with radius
$r_{\rm v} \sim 50$ Mpc/$h$ at $z\sim 0.1$ as shown in Fig.\ \ref{Tsimulation}, given 
the 1Gpc/$h$ simulation box-size, the projected 2D ISWRS temperature fluctuations are of the order of a 10$~\mu$K, while the ISWRS 
temperature from the void is at the order of 0.1 $\mu$K. To reach a 2$\sigma$ (3$\sigma$) detection, even without the primordial CMB 
fluctuations, the number of voids we would need to stack is $\sim$40000 (90000).
The fact that we have had a $\sim 2 \sigma$ signal (if ISW) with 470 voids is somewhat surprising, 
although again, there is a 5$\%$ chance of it being noise.
 
\subsection{Robustness Tests}
\label{tests}
A $\sim 2 \sigma$ signal is by no means significant. There is a 5$\%$ chance of it being random noise, in which case it may be 
sensitive to cuts we impose in the catalog. It is therefore instructive to test how 
the result varies with our selection criteria. To do this, we vary the void size cut around our fiducial choice over 
a wide range. Another quantity that could be used to remove unphysical voids is the ratio of 
the density on the ridge, $\rho_{\rm r}$, to its lowest density $\rho_{\rm min}$. A higher value of $\rho_{\rm r}/\rho_{\rm min}$ means
the void is more significant. This is a good quantifier of the probability $P$ of a void being real (one minus the probability of it arising in a Poisson process), 
\begin{equation}
P=1-\exp{[-5.12(\rho_{\rm r}/\rho_{\rm min}-1)-0.8(\rho_{\rm r}/\rho_{\rm min}-1)^{2.8}]} 
\label{eqn:poisson}
\end{equation}
\citep{Neyrinck08}. This ratio is also essentially the quantity known as persistence, a common measure of a feature's robustness in computational topology \citep[e.g.][]{Sousbie2011}.

In Fig.\ \ref{2Dparameter}, we show the mean CMB imprints of void samples characterized by different cuts in these two quantities.
The result from our fiducial cuts, as used in Fig.\ \ref{detection}, correspond to the pixel value at ($r_{\rm cut}/r_{\rm cut, default}=1$, $\rho_{\rm r}/\rho_{\rm min}=1$), \red{indicated by the red triangle}.
Pixels at larger $r_{\rm cut}/r_{\rm cut, default}$ represent results from stacking voids of larger radii, while those with 
larger $\rho_{\rm r}/\rho_{\rm min}$ are from stacking voids of greater significance. \red{Note that $r_{\rm cut}/r_{\rm cut, default}$ \red{constitutes an array of separate $r_{\rm cut}$ thresholds in each of the six 
sub-samples, since each has its own mean galaxy separation}. This is different from the case of treating voids of all the six sub-samples together, sorting them and thresholding them with one \red{single} value of $r_{\rm cut}$, as we do in Fig.\ \ref{detection}.} Overall, the stacked-filtered temperatures remain negative in this wide range of selection criteria. Moreover, there is an 
tendency for the amplitude to increase towards those pixels at the top-right corner, suggesting that the amplitude of the signal does 
increase with the size and the significance of voids, just like the ISW signal would be expected to behave. This suggests that the 
selection criteria we derive from our simulations are sensible in selecting physically meaningful voids that give cold CMB imprints 
in the observations.

We also try weighting voids by the probability that they are not spurious Poisson fluctuations, according to Eq.\ (\ref{eqn:poisson}). This reduces the parameter space to one dimension, 
$r_{\rm cut}/r_{\rm cut, default}$ (Fig.\ \ref{1Dparameter}). Overall, as we raise the criteria for 
$r_{\rm cut}/r_{\rm cut, default}$, the amplitude weighted-stacked-filtered temperature is increasing (with exceptions 
at $r_{\rm cut}/r_{\rm cut, default}\sim 1.5$) (left-hand panel of Fig.\ \ref{1Dparameter}), and the (effective) number of voids that passes the 
criteria also drops, as expected. As in Fig.\ \ref{2Dparameter}, This increase in signal with minimum void radius is similar to that seen in Fig.\ \ref{2Dparameter}, 
again suggesting that the putative signal is stable. The right-hand panel of Fig.\ \ref{1Dparameter} shows the dependence of the S/N on $r_{\rm cut}/r_{\rm cut, default}$. 
The S/N fluctuates from 0.7 to 2.5, and has a tendency to increase with $r_{\rm cut}/r_{\rm cut, default}$. Our fiducial cut \red{(as indicated by the red triangle)} happens to be at a local maximum, but not at the global one. 

\redd{
To see qualitatively our results shown in Fig.\ref{detection} may be affected by Poissonian confusion, we give one example of the cumulative stacked-filtered temperature weighted by the void probability, shown in the orange dotted curve
in Fig.\ref{detection}. The curve shows slightly smaller fluctuations than the orange solid curve when the number of voids used for
stacking is relatively small. In particular, a few large voids of lower probability have been down-weighted
(at $n_{\rm stack} \sim 100$) such that the stacked temperature is prevented from going positive, hence the amplitude of the
stacked-filtered temperature is increased. This suggests that the void-probability weighting scheme is helpful in reducing
Poisson fluctuations by down-weighting voids of low probability. However, as $n_{\rm stack}$ increases, or $r_{\rm v}$ drops,
weighting or no weighting makes little difference. The void-probability weighting scheme seems not efficient to help
increasing the signal. This may suggest that for relatively small voids, whether their filtered ISW temperatures are
negative or positive depends more strongly on their environments, i.e. void-in-cloud or void-in-void. The stacked
signal dependents weakly on void probabilities.}

We emphasize that results shown in Fig.\ \ref{2Dparameter} and Fig.\ \ref{1Dparameter} are merely a robustness test of our analysis. 
Seeing that there might be a larger signal for larger voids and more significant voids, we do not, however, change our the selection criteria 
and claim a higher-significance detection; this would constitute {\it a posteriori} bias.
\begin{figure}
\advance\leftskip -1.3cm
\scalebox{0.5}{
\includegraphics[angle=0]{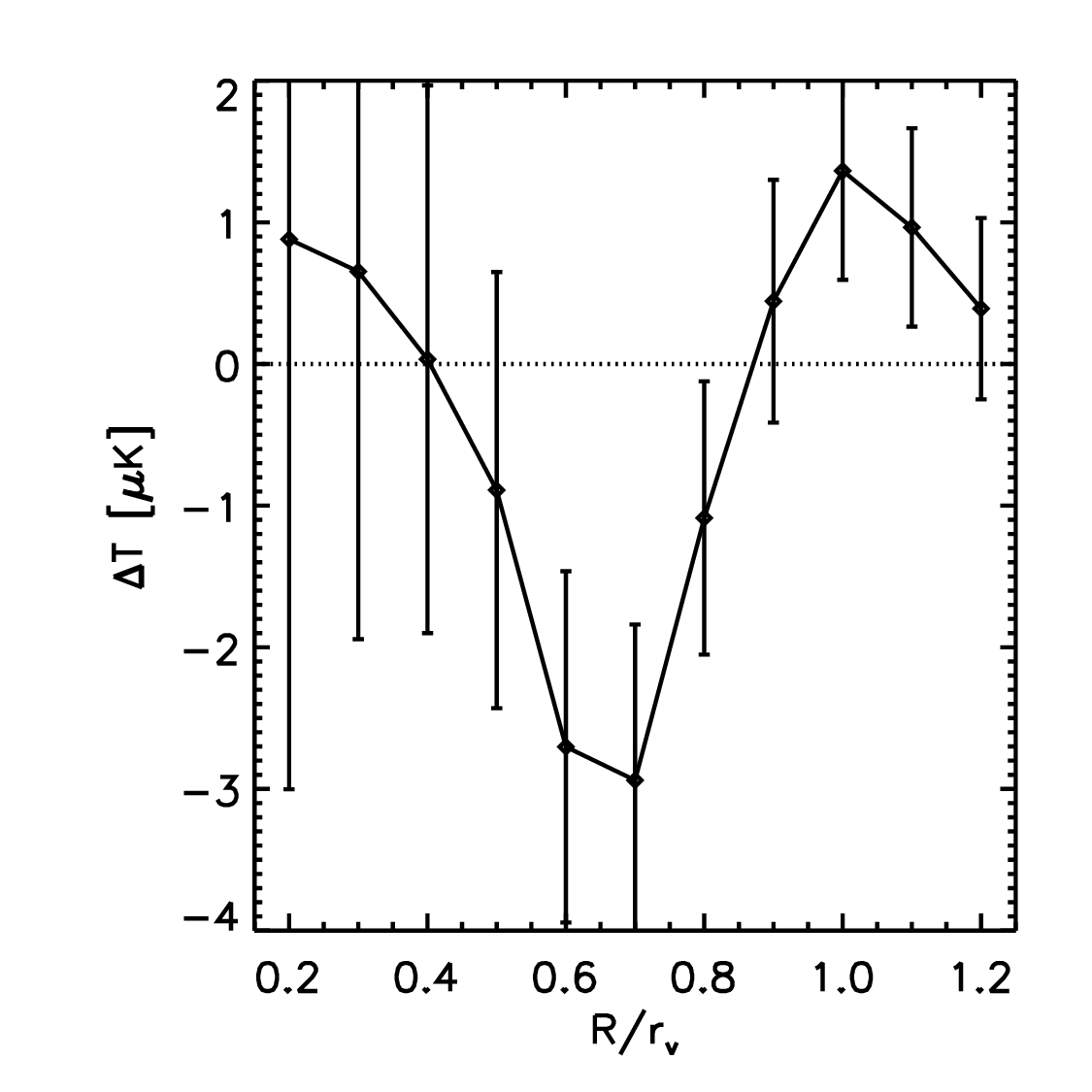}}
\caption{The observed, stacked void signal viewed through compensated top-hat filters of different radii $R$, relative to the void radius 
$r_{\rm v}$. It is to be compared with the red curve in Fig.\ \ref{Tsimulation}-B.}
\label{Fig.Tr}
\end{figure}

Finally, we also try varying the rescaling factor for the size of the filter radius. Results are shown in Fig.\ \ref{Fig.Tr}. 
Intriguingly, the same scaling factor of 0.6 that gives the largest amplitudes of the filtered ISWRS temperature in the simulations 
also gives nearly the largest signal in the data (0.7 in the data). This consistency with simulations 
regarding the optimal filter size is encouraging. It also seems to suggest that the measured signal may have 
similar profiles as the simulated ISWRS signal (when comparing Fig.\ \ref{Fig.Tr} to the red curve of Fig.\ \ref{Tsimulation}-B).
For these results, only results from the WMAP9-V band are shown for better illustration. The same tests with other WMAP bands 
and Planck SMICA give similar results.

\section{discussion and conclusions}
We have found that stacking CMB temperatures around the positions of voids from the SDSS DR7 spectroscopic galaxy catalog yields a 
temperature decrement between $2.6$ and $2.9 \mu$K, at about 2-$\sigma$ significance. When interpreted as the ISW and the Rees-Sciama effect, it 
is at odds with simulations of a $\Lambda$CDM universe at $\sim 2\sigma$. We want to emphasize our analysis is based strongly on simulations, 
which are very important in calibrating the raw data and methodology. In particular, 

A.) We have found a scaling factor for the compensated top-hat filter radius that optimizes the ISWRS detection. This factor is 0.6 times the effective void radius. In \citep{Ilic, Planck}, where similar analysis have been done independently with a similar void catalog, there is a strong indication 
that such a factor is needed to have the best S/N detection in the data. This suggests that there is indeed some (low-level) indication of an ISWRS signal in the data. 

B.) We emphasize two issues for void catalogues found using tracers of the density field: Poisson noise and environmental effects. 
There can be spurious voids from Poisson sampling. Also, voids live in different environments with 
different dynamical properties, and hence the ISWRS signal produced varies. A small void in an over-dense region that is contracting will correspond 
to an ISW hot spot greater than the size of the void. These issues are crucial to consider when pruning a void catalog for ISWRS analysis. 

C.) We use our simulations to clean up the `raw' void catalog, which is potentially noisy and unlikely to be optimal for ISW analysis. While our fiducial cuts on the void catalog might be conservative, the results seem to be robust when varying our thresholds around the 
fiducial ones. Interestingly, the signal increases if only the largest and most statistically significant voids are used, just as expected if the signal is the ISW effect.

Stacked CMB temperature maps using voids should suffer little contamination 
from other astrophysical processes, in that voids are relatively empty. 
The frequency independence of the result shown in the previous section suggests that it is unlikely to come from contamination
by radio sources or the thermal SZ effect \citep{SZ72}. It is also unlikely to be contaminated by the kSZ effect, which arises from the line-of-sight 
bulk motion of free electrons relative to the CMB. For the kSZ effect to induce the observed temperature decrement in voids, a void would have to have 
a high column density of free electrons, which is unlikely. 

Our measured average imprint somewhat bolsters the results of \citet[][G08]{Granett08}, and vice-versa.
The signals both go the same way, and are larger-amplitude than expected in a $\rm \Lambda$CDM universe. Our fiducial measured imprint is $\sim10$ 
times greater than expected in $\Lambda$CDM, although again we strongly caution that its statistical significance is only $\sim 2\sigma$.  The G08 statistical 
significance was 4.4$\sigma$ including both voids and clusters, and 3.7$\sigma$ including only the voids. Note that the present result is not entirely independent 
of G08, since they both use the CMB in the SDSS footprint, and there is a tiny overlap in redshift; this sample spans $z=0$-$0.44$, 
while the G08 sample spans $z=0.4$-$0.75$.

In principle, the tension between the detection of the ISWRS signal with the $\rm \Lambda$CDM model could be resolved by 
invoking non-Gaussianity. However the value of $f_{\rm NL}$ needed may be at the order $\sim1000$ \citep{HM12}, which is 
inconsistent with constraints from the CMB \citep[e.g.][]{Komatsu2011, PlanckNG}. Alternatively, voids in some models of 
modified gravity may grow larger and faster than in the $\rm \Lambda$CDM universe, giving a larger ISWRS signal \citet{Clampitt2012}. 
Whether or not such models can tolerate the high amplitude of the ISWRS signal while not violating other observational cosmological constraints 
remains an open question. Further investigation of this issue with surveys of larger volume and sky coverage is also needed to 
firmly confirm/resolve this tension. See an ISW study in the $f(R)$ modified gravity model in \citep{Cai2013}.

Throughout our paper, we have used our own algorithm to find voids in the SDSS volume limited samples. \citet{Sutter12} applied 
similar pipeline to the same SDSS DR7 data-sets, and constructed a public void catalog. We notice that while these two catalogues 
are similar in many aspects, there are subtle differences that may affect the ISW analysis. 
However, it is beyond the scope of our paper to understanding how the two catalogues differ. Nevertheless, we make our version of the void 
catalog public. The \citet{Sutter12} catalog has been used for the analysis of \citet{Ilic, Planck}. It is perhaps more valuable 
to have an relatively independent analysis. Also, \citet{Ilic, Planck} both have found some indications of 
a `2$\sigma$' signal by using all of the \citet{Sutter12} catalog at a filter scale radius of $\sim0.6$. In light of our calibration with 
simulations, this is exactly where the optimal signal is to be expected. Interestingly, with different catalogs and independent analysis 
pipelines, there are still suggestions of an ISWRS signal.

Finally, we note that although we have made significant progress toward optimal ISW analysis with voids through calibration with simulations, there is still room for improvement. Compensated filtering is essential, but the top-hat shape is likely not optimal; for example, the signal through a Mexican hat filter is more stable to changes in filter radius. Also, the optimal technique would likely involve an inverse-noise weighting for each void, which would involve ISWRS signals measured in simulations, as well as the expected `noise' from the primordial CMB. In this paper, we have kept our methodology close to that of previous works for comparison, but optimizing the method for ISWRS detection remains an interesting subject for future work.

\section*{Acknowledgments}
We thank P.\ M.\ Sutter for his helpful discussions and comments on an earlier version of 
the manuscript, \red{and S. Nadathur for useful comments on our later version. We thank the referee for a helpful report.} 
We thank Baojiu Li for providing the $N$-body simulations. MN is grateful for support from a New Frontiers grant from the 
Sir John Templeton Foundation. YC is supported by the Durham Junior Research Fellowship. 
IS acknowledges NASA grants NNX12AF83G and NNX10AD53G for support. YC, SC and CF acknowledge the support of the Science and Technology Facilities Council [Grant number ST/F001166/1]. 
Part of the calculations for this paper were performed on the ICC
Cosmology Machine, which is part of the DiRAC Facility jointly funded by STFC, the Large Facilities Capital Fund of BIS, and Durham University.
\red{Some of the results in this paper are derived using the {\sc HEALPix} package \citep{Gorski2005}}. 

\red{Funding for the SDSS and SDSS-II has been provided by the Alfred P. Sloan Foundation, the Participating Institutions, the National Science Foundation, the U.S. Department of Energy, the National Aeronautics and Space Administration, the Japanese Monbukagakusho, the Max Planck Society, and the Higher Education Funding Council for England. The SDSS Web Site is {\tt http://www.sdss.org/}.}

\red{The SDSS is managed by the Astrophysical Research Consortium for the Participating Institutions. The Participating Institutions are the American Museum of Natural History, Astrophysical Institute Potsdam, University of Basel, University of Cambridge, Case Western Reserve University, University of Chicago, Drexel University, Fermilab, the Institute for Advanced Study, the Japan Participation Group, Johns Hopkins University, the Joint Institute for Nuclear Astrophysics, the Kavli Institute for Particle Astrophysics and Cosmology, the Korean Scientist Group, the Chinese Academy of Sciences (LAMOST), Los Alamos National Laboratory, the Max-Planck-Institute for Astronomy (MPIA), the Max-Planck-Institute for Astrophysics (MPA), New Mexico State University, Ohio State University, University of Pittsburgh, University of Portsmouth, Princeton University, the United States Naval Observatory, and the University of Washington.}

\bibliography{ISW_Stacking}

\begin{thebibliography}{}
\expandafter\ifx\csname natexlab\endcsname\relax\def\natexlab#1{#1}\fi

\bibitem[{{Bennett} {et~al.}(2012){Bennett}, {Larson}, {Weiland}, {Jarosik},
  {Hinshaw}, {Odegard}, {Smith}, {Hill}, \& et~al.}]{WMAP9}
{Bennett}, C.~L., {Larson}, D., {Weiland}, J.~L., {et~al.} 2012, ArXiv
  e-prints, arXiv:1212.5225

\bibitem[{{Bielby} {et~al.}(2010){Bielby}, {Shanks}, {Sawangwit}, {Croom},
  {Ross}, \& {Wake}}]{Bielby2010}
{Bielby}, R., {Shanks}, T., {Sawangwit}, U., {et~al.} 2010, \mnras, 403, 1261

\bibitem[{{Cai} {et~al.}(2009){Cai}, {Cole}, {Jenkins}, \& {Frenk}}]{Cai09}
{Cai}, Y.-C., {Cole}, S., {Jenkins}, A., \& {Frenk}, C. 2009, \mnras, 396, 772

\bibitem[{{Cai} {et~al.}(2010){Cai}, {Cole}, {Jenkins}, \& {Frenk}}]{Cai10}
{Cai}, Y.-C., {Cole}, S., {Jenkins}, A., \& {Frenk}, C.~S. 2010, \mnras, 407,
  201

\bibitem[{{Cai} {et~al.}(2013){Cai}, {Li}, {Cole}, {Frenk}, \&
  {Neyrinck}}]{Cai2013}
{Cai}, Y.-C., {Li}, B., {Cole}, S., {Frenk}, C.~S., \& {Neyrinck}, M. 2013,
  ArXiv e-prints, arXiv:1310.6986

\bibitem[{{Ceccarelli} {et~al.}(2013){Ceccarelli}, {Paz}, {Lares}, {Padilla},
  \& {Lambas}}]{Ceccarelli2013}
{Ceccarelli}, L., {Paz}, D., {Lares}, M., {Padilla}, N., \& {Lambas}, D.~G.
  2013, \mnras, 434, 1435

\bibitem[{{Clampitt} {et~al.}(2013){Clampitt}, {Cai}, \& {Li}}]{Clampitt2012}
{Clampitt}, J., {Cai}, Y.-C., \& {Li}, B. 2013, \mnras, 431, 749

\bibitem[{{Crittenden} \& {Turok}(1996)}]{CrittendenTurok96}
{Crittenden}, R.~G., \& {Turok}, N. 1996, Physical Review Letters, 76, 575

\bibitem[{{Davis} {et~al.}(1985){Davis}, {Efstathiou}, {Frenk}, \&
  {White}}]{Davis1985}
{Davis}, M., {Efstathiou}, G., {Frenk}, C.~S., \& {White}, S.~D.~M. 1985, \apj,
  292, 371

\bibitem[{{Eisenstein} {et~al.}(2001){Eisenstein}, {Annis}, {Gunn}, {Szalay},
  {Connolly}, {Nichol}, {Bahcall}, {Bernardi}, {Burles}, {Castander},
  {Fukugita}, {Hogg}, {Ivezi{\'c}}, {Knapp}, {Lupton}, {Narayanan}, {Postman},
  {Reichart}, {Richmond}, {Schneider}, {Schlegel}, {Strauss}, {SubbaRao},
  {Tucker}, {Vanden Berk}, {Vogeley}, {Weinberg}, \& {Yanny}}]{Eisenstein2001}
{Eisenstein}, D.~J., {Annis}, J., {Gunn}, J.~E., {et~al.} 2001, \aj, 122, 2267

\bibitem[{{Flender} {et~al.}(2012){Flender}, {Hotchkiss}, \&
  {Nadathur}}]{Flender12}
{Flender}, S., {Hotchkiss}, S., \& {Nadathur}, S. 2012, ArXiv e-prints,
  arXiv:1212.0776

\bibitem[{{Giannantonio} {et~al.}(2012){Giannantonio}, {Crittenden}, {Nichol},
  \& {Ross}}]{Giannatonio12}
{Giannantonio}, T., {Crittenden}, R., {Nichol}, R., \& {Ross}, A.~J. 2012,
  \mnras, 426, 2581

\bibitem[{{Giannantonio} {et~al.}(2008){Giannantonio}, {Scranton},
  {Crittenden}, {Nichol}, {Boughn}, {Myers}, \& {Richards}}]{Giannatonio08}
{Giannantonio}, T., {Scranton}, R., {Crittenden}, R.~G., {et~al.} 2008, \prd,
  77, 123520

\bibitem[{{G{\'o}rski} {et~al.}(2005){G{\'o}rski}, {Hivon}, {Banday},
  {Wandelt}, {Hansen}, {Reinecke}, \& {Bartelmann}}]{Gorski2005}
{G{\'o}rski}, K.~M., {Hivon}, E., {Banday}, A.~J., {et~al.} 2005, \apj, 622,
  759

\bibitem[{{Granett} {et~al.}(2008){Granett}, {Neyrinck}, \&
  {Szapudi}}]{Granett08}
{Granett}, B.~R., {Neyrinck}, M.~C., \& {Szapudi}, I. 2008, \apjl, 683, L99

\bibitem[{{Hamaus} {et~al.}(2013){Hamaus}, {Wandelt}, {Sutter}, {Lavaux}, \&
  {Warren}}]{HamausEtal2013}
{Hamaus}, N., {Wandelt}, B.~D., {Sutter}, P.~M., {Lavaux}, G., \& {Warren},
  M.~S. 2013, ArXiv e-prints, arXiv:1307.2571

\bibitem[{{Hern{\'a}ndez-Monteagudo}(2010)}]{HM2010}
{Hern{\'a}ndez-Monteagudo}, C. 2010, \aap, 520, A101

\bibitem[{{Hernandez-Monteagudo} \& {Smith}(2012)}]{HM12}
{Hernandez-Monteagudo}, C., \& {Smith}, R.~E. 2012, MNRAS, submitted,
  arXiv:1212.1174

\bibitem[{{Ho} {et~al.}(2008){Ho}, {Hirata}, {Padmanabhan}, {Seljak}, \&
  {Bahcall}}]{Ho08}
{Ho}, S., {Hirata}, C., {Padmanabhan}, N., {Seljak}, U., \& {Bahcall}, N. 2008,
  \prd, 78, 043519

\bibitem[{{Ili{\'c}} {et~al.}(2013){Ili{\'c}}, {Langer}, \& {Douspis}}]{Ilic}
{Ili{\'c}}, S., {Langer}, M., \& {Douspis}, M. 2013, \aap, 556, A51

\bibitem[{{Komatsu} {et~al.}(2011){Komatsu}, {Smith}, {Dunkley}, {Bennett},
  {Gold}, {Hinshaw}, {Jarosik}, {Larson}, \& et~al.}]{Komatsu2011}
{Komatsu}, E., {Smith}, K.~M., {Dunkley}, J., {et~al.} 2011, \apjs, 192, 18

\bibitem[{{Larson} {et~al.}(2011){Larson}, {Dunkley}, {Hinshaw}, {Komatsu},
  {Nolta}, {Bennett}, {Gold}, {Halpern}, \& et~al.}]{Larson11}
{Larson}, D., {Dunkley}, J., {Hinshaw}, G., {et~al.} 2011, \apjs, 192, 16

\bibitem[{{Li} {et~al.}(2013){Li}, {Hellwing}, {Koyama}, {Zhao}, {Jennings}, \&
  {Baugh}}]{Li2013}
{Li}, B., {Hellwing}, W.~A., {Koyama}, K., {et~al.} 2013, \mnras, 428, 743

\bibitem[{{L{\'o}pez-Corredoira} {et~al.}(2010){L{\'o}pez-Corredoira}, {Sylos
  Labini}, \& {Betancort-Rijo}}]{Lopez-Corredoira2010}
{L{\'o}pez-Corredoira}, M., {Sylos Labini}, F., \& {Betancort-Rijo}, J. 2010,
  \aap, 513, A3

\bibitem[{{Nadathur} \& {Hotchkiss}(2013)}]{Nadathur2013}
{Nadathur}, S., \& {Hotchkiss}, S. 2013, ArXiv e-prints, arXiv:1310.2791

\bibitem[{{Nadathur} {et~al.}(2012{\natexlab{a}}){Nadathur}, {Hotchkiss}, \&
  {Sarkar}}]{Nadathur12}
{Nadathur}, S., {Hotchkiss}, S., \& {Sarkar}, S. 2012{\natexlab{a}}, JCAP, 6,
  42

\bibitem[{{Nadathur} {et~al.}(2012{\natexlab{b}}){Nadathur}, {Hotchkiss}, \&
  {Sarkar}}]{Nadathur2012}
---. 2012{\natexlab{b}}, JCAP, 6, 42

\bibitem[{{Neyrinck}(2008)}]{Neyrinck08}
{Neyrinck}, M.~C. 2008, \mnras, 386, 2101

\bibitem[{{Neyrinck} {et~al.}(2005){Neyrinck}, {Gnedin}, \&
  {Hamilton}}]{NeyrinckEtal2005}
{Neyrinck}, M.~C., {Gnedin}, N.~Y., \& {Hamilton}, A.~J.~S. 2005, \mnras, 356,
  1222

\bibitem[{{P{\'a}pai} \& {Szapudi}(2010)}]{PapaiSzapudi2010}
{P{\'a}pai}, P., \& {Szapudi}, I. 2010, \apj, 725, 2078

\bibitem[{{P{\'a}pai} {et~al.}(2011){P{\'a}pai}, {Szapudi}, \&
  {Granett}}]{PapaiEtal2011}
{P{\'a}pai}, P., {Szapudi}, I., \& {Granett}, B.~R. 2011, \apj, 732, 27

\bibitem[{{Planck Collaboration} {et~al.}(2013{\natexlab{a}}){Planck
  Collaboration}, {Ade}, {Aghanim}, {Armitage-Caplan}, {Arnaud}, {Ashdown},
  {Atrio-Barandela}, {Aumont}, {Baccigalupi}, {Banday}, \& et~al.}]{PlanckXII}
{Planck Collaboration}, {Ade}, P.~A.~R., {Aghanim}, N., {et~al.}
  2013{\natexlab{a}}, ArXiv e-prints, arXiv:1303.5072

\bibitem[{{Planck Collaboration} {et~al.}(2013{\natexlab{b}}){Planck
  Collaboration}, {Ade}, {Aghanim}, {Armitage-Caplan}, {Arnaud}, {Ashdown},
  {Atrio-Barandela}, {Aumont}, {Baccigalupi}, {Banday}, \& et~al.}]{Planck}
---. 2013{\natexlab{b}}, ArXiv e-prints, arXiv:1303.5079

\bibitem[{{Planck Collaboration} {et~al.}(2013{\natexlab{c}}){Planck
  Collaboration}, {Ade}, {Aghanim}, {Armitage-Caplan}, {Arnaud}, {Ashdown},
  {Atrio-Barandela}, {Aumont}, {Baccigalupi}, {Banday}, \& et~al.}]{PlanckNG}
---. 2013{\natexlab{c}}, ArXiv e-prints, arXiv:1303.5084

\bibitem[{{Platen} {et~al.}(2007){Platen}, {van de Weygaert}, \&
  {Jones}}]{PlatenEtal2007}
{Platen}, E., {van de Weygaert}, R., \& {Jones}, B.~J.~T. 2007, \mnras, 380,
  551

\bibitem[{{Rassat} {et~al.}(2007){Rassat}, {Land}, {Lahav}, \&
  {Abdalla}}]{Rassat2007}
{Rassat}, A., {Land}, K., {Lahav}, O., \& {Abdalla}, F.~B. 2007, \mnras, 377,
  1085

\bibitem[{{Rees} \& {Sciama}(1968)}]{Rees68}
{Rees}, M.~J., \& {Sciama}, D.~W. 1968, \nat, 217, 511

\bibitem[{{Sachs} \& {Wolfe}(1967)}]{Sachs67}
{Sachs}, R.~K., \& {Wolfe}, A.~M. 1967, \apj, 147, 73

\bibitem[{{Sawangwit} {et~al.}(2010){Sawangwit}, {Shanks}, {Cannon}, {Croom},
  {Ross}, \& {Wake}}]{Sawangwit10}
{Sawangwit}, U., {Shanks}, T., {Cannon}, R.~D., {et~al.} 2010, \mnras, 402,
  2228

\bibitem[{{Schaap} \& {van de Weygaert}(2000)}]{SchaapVdw2000}
{Schaap}, W.~E., \& {van de Weygaert}, R. 2000, \aap, 363, L29

\bibitem[{{Seljak}(1996)}]{Seljak96}
{Seljak}, U. 1996, \apj, 460, 549

\bibitem[{{Sheth} \& {van de Weygaert}(2004)}]{ShethVDWeygaert2004}
{Sheth}, R.~K., \& {van de Weygaert}, R. 2004, \mnras, 350, 517

\bibitem[{{Smith} {et~al.}(2009){Smith}, {Hern{\'a}ndez-Monteagudo}, \&
  {Seljak}}]{Smith2009}
{Smith}, R.~E., {Hern{\'a}ndez-Monteagudo}, C., \& {Seljak}, U. 2009, \prd, 80,
  063528

\bibitem[{{Sousbie}(2011)}]{Sousbie2011}
{Sousbie}, T. 2011, \mnras, 414, 350

\bibitem[{{Strauss} {et~al.}(2002){Strauss}, {Weinberg}, {Lupton}, {Narayanan},
  {Annis}, {Bernardi}, {Blanton}, {Burles}, {Connolly}, {Dalcanton}, {Doi},
  {Eisenstein}, {Frieman}, {Fukugita}, {Gunn}, {Ivezi{\'c}}, {Kent}, {Kim},
  {Knapp}, {Kron}, {Munn}, {Newberg}, {Nichol}, {Okamura}, {Quinn}, {Richmond},
  {Schlegel}, {Shimasaku}, {SubbaRao}, {Szalay}, {Vanden Berk}, {Vogeley},
  {Yanny}, {Yasuda}, {York}, \& {Zehavi}}]{Strauss2002}
{Strauss}, M.~A., {Weinberg}, D.~H., {Lupton}, R.~H., {et~al.} 2002, \aj, 124,
  1810

\bibitem[{{Sunyaev} \& {Zeldovich}(1972)}]{SZ72}
{Sunyaev}, R.~A., \& {Zeldovich}, Y.~B. 1972, Comments on Astrophysics and
  Space Physics, 4, 173

\bibitem[{{Sutter} {et~al.}(2012){Sutter}, {Lavaux}, {Wandelt}, \&
  {Weinberg}}]{Sutter12}
{Sutter}, P.~M., {Lavaux}, G., {Wandelt}, B.~D., \& {Weinberg}, D.~H. 2012,
  \apj, 761, 44

\bibitem[{{Sutter} {et~al.}(2013){Sutter}, {Lavaux}, {Wandelt}, \&
  {Weinberg}}]{SutterEtal2013b}
---. 2013, ArXiv e-prints, arXiv:1310.5067

\bibitem[{{Watson} {et~al.}(2013){Watson}, {Diego}, {Gottl{\"o}ber}, {Iliev},
  {Knebe}, {Mart{\'{\i}}nez-Gonz{\'a}lez}, {Yepes}, {Barreiro},
  {Gonz{\'a}lez-Nuevo}, {Hotchkiss}, {Marcos-Caballero}, {Nadathur}, {Vielva},
  \& {.}}]{Watson2013}
{Watson}, W.~A., {Diego}, J.~M., {Gottl{\"o}ber}, S., {et~al.} 2013, ArXiv
  e-prints, arXiv:1307.1712

\bibitem[{{Zhang} \& {Huterer}(2010)}]{Zhang2010}
{Zhang}, R., \& {Huterer}, D. 2010, Astroparticle Physics, 33, 69

\end{thebibliography}
\bibliographystyle{apj}

\end{document}